\documentclass[prd,twocolumn,showpacs,eqsecnum]{revtex4}
\usepackage[dvips]{graphicx}
\usepackage{color}
\usepackage{amsmath}
\usepackage{amssymb}
\begin{document}
%
%
\preprint{LA-UR 04-460}
\title
   { Renormalizing the Schwinger-Dyson equations in the
   Auxiliary Field Formulation of $\lambda \phi^4$ Field Theory
   }

\author{Fred~Cooper}
\email{cooper@nsf.gov}
\affiliation{National Science Foundation,
   Division of Physics,
   Arlington, VA 22230}
\affiliation{Santa Fe Institute,
   Santa Fe, NM 87501}
\affiliation{Theoretical Division,
   Los Alamos National Laboratory,
   Los Alamos, NM 87545}
\author{Bogdan~Mihaila}
\email{bmihaila@lanl.gov} \affiliation{Theoretical Division,
   Los Alamos National Laboratory,
   Los Alamos, NM 87545}
\author{John~F.~Dawson}
\email{john.dawson@unh.edu}
\affiliation{
   Department of Physics,
   University of New Hampshire,
   Durham, NH 03824}

\date{\today}
\begin{abstract}
In this paper we study the renormalization of the Schwinger-Dyson
equations that arise in the auxiliary field formulation of the
O(N) $\phi^4$ field theory.
The auxiliary field formulation allows a simple interpretation of
the large-N expansion as a loop expansion of the generating
functional in the auxiliary field $\chi$, once the effective
action is obtained by integrating over the $\phi$ fields. Our all
orders result is then used to obtain finite renormalized
Schwinger-Dyson equations  based on truncation expansions which
utilize the two-particle irreducible (2-PI) generating function
formalism. We first do an all orders renormalization of the two-
and three-point function equations in the vacuum  sector. This
result is then used  to obtain explicitly finite and
renormalization constant independent self-consistent S-D equations
valid to order~1/N, in both 2+1 and 3+1 dimensions.  We compare
the results for the real and imaginary parts of the renormalized
Green's functions with the related \emph{sunset} approximation to
the 2-PI equations discussed by Van Hees and Knoll, and comment on
the importance of the Landau pole effect.

\end{abstract}
%
\maketitle

%
%
\section{Introduction}
\label{sec:intro}

Recently there has been interest in studying field theory using
two-particle irreducible (2-PI) methods~\cite{ref:2PI} in both
finite temperature~\cite{ref:temp, ref:VanHees, ref:sunset} and
non-equilibrium situations~\cite{ref:Hu, ref:noneq1, ref:noneq2,
berges-2PI, 4d_1, 4d_2, 3d_1, baacke, juchem, aarts}. The value of
the 2-PI formalism for non-equilibrium problems is that it allows
one to make approximations that go beyond the Hartree or large-N
approximation without encountering the serious problems of
secularity found in a straightforward expansion about the Hartree
or leading-order large-N approximation using the generating
functional or, equivalently, the one-particle irreducible (1-PI)
action~\cite{ref:sec}.  The 2-PI methods lead to self-consistent
equations for the Green's functions which require non-perturbative
renormalization. Recently the renormalization of the equations
obtained from the 2-PI approach applied to the standard
formulation of $\phi^4$ field theory
(first discussed  by Calzetta and Hu~\cite{ref:Calzetta}) has been
considered by both Van Hees and
Knoll~\cite{ref:VanHees,ref:sunset}, and by Blaizot et.
al.~\cite{ref:Blaizot}. This direct loop expansion of $\phi^4$
field theory  is a summation of the coupling constant expansion
and needs to be resummed in order to be related to a~1/N
expansion~\cite{ref:noneq2}. The approach to renormalization in
the above works was based on formal
Bogoliubov-Parasiuk-Hepp-Zimmermann (BPHZ)~\cite{ref:BPHZ} and
dimensional regularization~\cite{dreg} methods, rather than
multiplicative renormalization~\cite{bd2, Haymaker1}.  The
advantage of the multiplicative renormalization approach for
initial-value problems is that it lends itself more easily to the
momentum space cutoffs that occur when one uses  numerical methods
to solve the integro-differential equations of the closed time
path (CTP) formalism~\cite{CTP}.  In non-equilibrium situations,
the Green's function equations are usually only spatially
translational invariant and to make the calculational tractable, a
maximum 3-momentum is introduced (3-momentum cut-off $\Lambda$).
In dynamical situations the calculational schemes that have been
used usually rely on mode expansions for the quantum fields which
introduce non-covariant momentum cutoffs. Therefore, for practical
reasons it is useful to consider direct renormalization of the
Schwinger-Dyson (S-D) equations that are rendered finite by
momentum space cutoffs. Such an approach was very useful in the
leading order in large-N approximation, where we found that using
lattice versions of the renormalization scheme gave us results (as
long as we were far from the Landau pole) that became independent
of the cutoff  for a
wide range of cutoffs 
when we kept renormalized parameters fixed~\cite{CHKMPA,dcc}.  It
was also important when studying the time evolution of the
(expectation value of the)  energy-momentum tensor to understand
the non-covariant nature of the cutoff scheme, so that the correct
physical energy densities and pressures could be extracted from
the (non-covariant) situation arising in the truncation scheme
used for numerical simulation~\cite{ref:tmunu}.  By not
automatically subtracting off the logarithmical ($\log$)
divergences related to coupling constant renormalization, we were
able to have another check on the numerical simulations by
studying  how the simulations became independent of the cutoff for
fixed renormalized parameters.

In any truncation scheme, such as expanding the 2-PI generating
functional in terms of loops or 1/N, the renormalization has to be
guided by the structure of the {\it exact} renormalized S-D
equations.  Thus as a preliminary step to renormalizing the
truncated S-D equations, one needs to know the structure of the
{\it exact} renormalized S-D equations.   The strategy for
obtaining the renormalized S-D equations and then using them to
renormalize the next to leading order in~1/N is discussed
in~\cite{Haymaker1, CHKMPA}. In those papers, however, the
perturbative~1/N expansion was discussed and {\it not} the
resummed~1/N expansion obtained from the 2-PI formalism.

Our procedure is as follows: First we derive the exact S-D
equations for the auxiliary field formalism. These S-D equations
are simpler than those of the original formulation of $\phi^4$
field theory because the only quantities that need renormalization
are the propagators for the field $\phi$ and the auxiliary field
$\chi$, as well as the three-particle  $\chi \phi \phi$
irreducible vertex. Analyzing graphs one finds that once those
quantities are renormalized one never generates a new divergence
in the coupling of four $\phi$ particles. In the cutoff S-D
approach to renormalization, one has to show not only that the
renormalized equations are finite, but also that they are
independent of all the (infinite) renormalization constants. That
is, all the equations need to be written in terms of the
renormalized Green's functions, vertices and masses. The reason
for using the auxiliary field formulation of $\lambda \phi^4$
field theory is that the~1/N expansion has a simple interpretation
as a loop expansion [in the auxiliary field] of the generating
functional of the effective action obtained by integrating out the
scalar fields keeping $\chi$ constant~\cite{on1,root}. The S-D
equations arising from this auxiliary field formulation was first
discussed in Ref.~\cite{bcg}.

The~1/N expansion is an asymmetric expansion which treats the
$\phi$ field exactly (with $\chi$ fixed), and then counts loops in
$\chi$.  Thus the 2-PI formalism, which treats $\phi$ and $\chi$
on an equal footing, is not a natural   formalism for
incorporating the large-N approximation except in leading order.
Its main virtue is that it leads to self-consistent approximations
that are energy conserving and non-secular when applied in
non-equilibrium contexts.  The basic propagators that occur in the
large-N expansion, when viewed as propagators coming from the
effective Lagrangian obtained after integrating out the $\phi$
field in the path integral,  have different behavior with regard
to~1/N~\cite{berges-2PI}: the $\phi \phi$ propagator $G$  being
$\mathcal{O}(1)$, the $ \chi \chi$ propagator $D$ being of
order~1/N and the $\chi \phi$ mixed propagator  $K$ (which
vanishes if symmetry is unbroken) is also of order~1/N. Thus the
first nontrivial two-loop 2-PI vacuum graph  has terms $GGD
\sim$~1/N and $KKG \sim$~1/N$^2$.  At the level of the equation
for the inverse two-point Green's functions, the first nontrivial
approximation (counting loops) has no {\it vertex} corrections,
but mixes orders of~1/N.   Two approximations which have been
studied recently in studies of thermalization have been based on
keeping one or both of these two-loop
graphs~\cite{ref:noneq1,ref:noneq2}. The first approximation has
been called the 2-PI~1/N expansion and the second the bare-vertex
approximation (BVA). Both these approximations are identical when
$\langle \phi \rangle =0$ and thus the renormalization scheme is
identical for both approximations.  We find that both
approximations after renormalization require that the {\it
renormalized} vertex function is set to 1. Thus the BVA is a
misnomer in dimensions greater  than $2+1$ where wave function and
vertex function renormalizations are necessary.

Recently, the 2-PI~1/N expansion has been used to study the
nonequilibrium dynamics of field theory in 3+1 dimensions: First,
in Ref.~\cite{4d_1}, the 2-PI~1/N was used to study the parametric
resonance of an O(N) symmetric scalar theory, at very weak
coupling constant. Secondly, in Ref~\cite{4d_2}, the same
approximation was used to investigate the nonequilibrium dynamics
of a 3+1 dimensional theory with Dirac fermions coupled to scalars
via a chirally invariant Yukawa interaction. In the later case,
the system was shown to reach, at late times, a state which was
well described by a thermal distribution. In the above work, only
the quadratic mass divergences were eliminated, and the
renormalized mass extracted from the oscillation frequency.
Determining the renormalized mass dynamically makes it a bit more
difficult to study the behavior of the theory for fixed
renormalized mass as a function of the momentum space cutoff. The
advantage of having renormalized S-D equations which depend on a
given fixed renormalized mass is two-fold.  In the static case it
allows for a rapidly convergent iteration method, since the
influence of cuts on the Green's function start only at $p^2 = 9
m^2$, where $m$ is the physical mass.  Secondly, in the time
evolution problem, the range of cutoffs which lead to results
insensitive to the cutoff and the Landau pole value can be easily
explored. Another recent development is the use of the 2-PI
methods to obtain universal behavior at a critical point~\cite
{3d_1}, where encouraging results for critical indices have been
obtained.

The S-D equations we investigate here were studied earlier by
Bender, Cooper and Guralnik~\cite{bcg,CHKMPA} and  are similar in
structure to those obtained for the Gross-Neveu model and analyzed
at all orders by Haymaker, Cooper et
al.~{\cite{Haymaker1}.  Our approach to the
renormalization of the S-D equations  parallels the treatment in
that body of work and allows a simple renormalization scheme at
order~1/N where the renormalized vertex is replaced by~$1$.

We organize this paper as follows.  In Section~\ref{sec:basics} we
introduce the auxiliary field formalism and discuss the~1/N
expansion as well as derive the unrenormalized S-D equations for
the two- and three-point functions. We also discuss 2-PI expansion
and the BVA. In Section~\ref{sec:allorders} we discuss the
renormalization of the vacuum sector of the unbroken theory.  In
Section~\ref{sec:1overN} we display the self-consistent
renormalized S-D equations for the vacuum sector valid to order
1/N. We solve these equations in the vacuum sector by an iteration
scheme based on utilizing the lowest order in 1/N results and
dispersion relations using a scheme used in
Ref.~\cite{ref:sunset}. We compare our results to the related
sunset approximation discussed in Ref.~\cite{ref:sunset} and find
that at large coupling constant, $g$, there are significant
differences between the sunset approximation and the
next-to-leading order in $1/N$ self-consistent approximation.
Finally, in Section~\ref{sec:landau}, we comment on the effect of
the Landau Pole (triviality of continuum $\lambda \phi^4 $ field
theory~\cite{phi4}) on our treatment of the 3+1 dimensional problem.

%
%
\section{Auxiliary Field Formulation of  the O(N) model}
\label{sec:basics}

Consider the Lagrangian for O(N) symmetry:
\begin{align}
   \mathcal{L}[\phi,\partial_\mu \phi]
   \ = \ &
   \frac{1}{2} \,
      \Bigl [
         \partial_\mu \phi_i(x) \, \partial^\mu \phi_i(x)
         +
         \mu^2 \, \phi_i^2(x)
      \Bigr ]
   \notag \\ &
   -
   \frac{g}{8} \, [ \phi_i^2(x) ]^2
   -
   \frac{\mu^4 }{2 g} \>.
\end{align}
Here, $g$ denotes the scaled coupling constant
$g=\lambda/N$. The Einstein summation convention for repeated
indices is implied throughout this paper.

We introduce a composite field $\chi(x)$ by adding to the
Lagrangian a term
\begin{equation}
   + \
   \frac{1}{2 g} \,
   \left \{
      \chi(x) - \frac{g}{2} \,
      \Bigl [ \, \phi_i^2(x) - 
                 \frac{2 \mu^2}{g} \, \Bigr ]
   \right \}^2
   \>.
\end{equation}
This gives a Lagrangian of the form
\begin{align}
   \mathcal{L}[\phi, \chi, \partial_\mu \phi]
   \ = \ &
   \frac{1}{2} \,
      \Bigl [
         \partial_\mu \phi_i(x) \, \partial^\mu \phi_i(x)
         - \chi(x) \, \phi_i^2(x)
      \Bigr ]
   \notag \\ &
   +
   \frac{ \mu^2 \, \chi(x) }{g}
   +
   \frac{ \chi^2(x) }{ 2 \, g }  \>,
\end{align}
which leads to the classical equations of motion
\begin{gather}
   \Bigl  [ \Box +  \chi(x) \Bigr ] \, \phi_i(x)
   = 0 \>,
\end{gather}
and the constraint (``gap'') equation
\begin{gather}
   \chi(x)
   =
   - \mu^2
   +
   \frac{g}{2} \, \sum_i \, \phi_i^2(x)
   \>.
\end{gather}

%
%
\subsection{The Large-N expansion}

The generating functional for the graphs of the auxiliary field
formalism is given by
\begin{align}
   Z[j_i, K] \ = \ &
   \exp( i \, N \, W[j,K])
   \\ \notag
   \ = \ &
   \int \mathrm{d} \chi \prod_{i=1}^{N} \mathrm{d} \phi_i
      \exp \biggl\{ i \int \mathrm{d} x
      \Bigl [ \mathcal{L} +  j_i(x) \phi_i(x)
   \\ \notag & \qquad \qquad
   + i N K(x)
   \chi(x)
   \Bigr ] \biggr \}
   \>.
\end{align}
The large-N expansion is obtained by integrating the Gaussian path
integrals for $\phi_i$, letting each $j_i = j$, and setting the
free inverse propagator $G_{0\, ij} = \delta_{ij} \, G_0$
(see~\cite{on1, root, bcg,CHKMPA}). This results in an effective
action
\begin{align}
   S_{eff}[\chi; j, K] / N = &
   \int \mathrm{d} x \, \Bigl [  \frac{ \mu^2 \, \chi(x) }{\lambda}
   +
   \frac{ \chi^2(x) }{ 2 \, \lambda } + K(x) \chi(x)
   \Bigr ]
   \notag \\ &
   + \frac{1}{2} \ j \circ G \circ j
   + \frac{i}{2} \ {\rm Tr} \ln G_0^{-1}[\chi]
   \>,
\end{align}
where
\begin{equation}
   G_0^{-1}[\chi](x-y] \ = \
   \Bigl [ \Box+ \chi(x) \Bigr ] \, \delta(x-y)
   \>,
\end{equation}
and we have introduced the notation
\begin{equation}
   \int \mathrm{d}x \int \mathrm{d}y \ j_i (x) \ G[\chi]_{ij} (x,y) \ j_j(y)
   = N \
   j \circ G \circ j
   \>.
\label{comp}
\end{equation}

The evaluation of the remaining path integral for $\chi$ by
steepest descent then leads to the~1/N expansion. The stationary
phase-point of the integrand $\chi_s[j,K]$ is determined
(implicitly) by the relation,
\begin{align}
   K(x)
   +
   \biggl \{ &
      {\frac {1} {\lambda}}\, \Bigl [ \chi (x) + \mu^2 \Bigr]
      -       \frac{1}{2} \ j \circ G(\ ,x) G (x,\ )\circ j
   \notag \\ &
      +       \frac{i}{2} \ G_0(x,x)\biggr \}_{\chi = \chi_s} = 0
   \>. \label{chis}
\end{align}
Keeping only the Gaussian fluctuations we obtain for the $W$
\begin{align}
   W[j,K]
   \equiv &
   W^{(0)}
   + {\frac {1} {N}} W^{(1)}
\label{WJK}
   \\ \notag
   = &
   \frac{1}{\lambda} \chi_s \circ (\frac{\chi_s}{2} + \mu^2)
   +  K  \circ \chi_s
   + {\frac{1}{2}} \, j \circ G_0 [\chi_s]\circ j
   \\ \notag &
   + {\frac{i}{2}} \, {\rm Tr} \ln G_0^{-1}[\chi_s]
   + {\frac{i}{2N}} \, {\rm Tr} \ln D_0^{-1}[j,K]
\end{align}
where $\chi_s$ is to be viewed as a function of the sources $j$
and $K$ through Eq.~\eqref{chis} above, and order~1/N$^2$ terms
have been dropped.  The bare inverse propagator for the auxiliary
field $\chi$, $D_0$, is defined as the second derivative of the
effective action with respect to $\chi$  at the stationary phase
point and its value is:
\begin{align}
   D^{-1}[j,K](x,y) \equiv &
   - \frac {1} {g} \, \delta^4(x,y)
   \\ \notag &
   -
   N \
   \Bigl[ j \circ G_0(\ ,x)G_0(x,y)G_0(y,\ )\circ j
   \\ \notag & \qquad \qquad
   - \frac{i}{2} \, G_0(x,y) G_0(y,x)\Bigr]_{\chi = \chi_s}
   \>.
\end{align}

The perturbative~1/N expansion  for the connected Green's
functions is obtained by treating all terms beyond the Gaussian
term in the effective action perturbatively and is equivalent to a
loop expansion in $\chi$ for the effective action.  Unfortunately
this expansion has the same defect as ordinary perturbation theory
when applied to time evolution problems in that it displays
secular behavior as demonstrated in~\cite{ref:sec} beyond the
leading order in large-N.  It is precisely for this reason that
the 2-PI approach has proven so useful, since it leads to
self-consistent S-D equation approximations that seem to be free
from secularity.

For completeness, we note that the generating functional for the
1-PI graphs, which is usually called the effective action, is the
Legendre transform of $W[j,K]$ to the new variables which are the
expectation value of the fields $\phi ={\delta W}/{\delta j}$, and
$\chi = {\delta W}/{\delta K}$. That is
\begin{equation}
\Gamma  [\phi, \chi ] /N \equiv W - j\circ \phi-  K \circ \chi ,
\label{Sdef}
\end{equation}
To order~1/N one obtains
\begin{equation}
\Gamma [\phi, \chi]/N  = S_{cl}[\phi,\chi]   + {\frac{i}{2}} {\rm
Tr} \ln G_0^{-1}[\chi] + {\frac{i}{2N}}  {\rm Tr} \ln
D_0^{-1}[\phi \chi] \>, \label{Seff}
\end{equation}
where
\begin{align}
   D_0^{-1}[\phi, \chi](x,y) = &
   -  \frac{1}{g} \ \delta^4(x,y)
   - N \, \phi(x)G_0[\chi](x,y) \phi(y)
   \notag \\ &
   + {\frac{iN}{2}} \, G_0[\chi](x,y) G[\chi](y,x)
   \>.
 \label{Dinv}
 \end{align}

%
%
\subsection{The S-D equations}

The  S-D equations in the auxiliary field formulation treats the
fields $\phi$ and $\chi$ and their two-point correlation functions
on an equal basis. Thus for considering the S-D equations or the
2-PI effective action, it is useful to use the extended field and
extended current notations,
\begin{align}
   \phi_\alpha(x)
   &=
   [ \chi(x), \phi_1(x), \phi_2(x), \ldots , \phi_N(x)] \>,
   \label{eq:xjextended}
   \\ \notag
   j_\alpha(x)
   &=
   [J(x), j_1(x), j_2(x), \ldots , j_N(x)]  \>, \
   \alpha=0,1,\ldots,N
   \>,
\end{align}
where $J(x) = N K(x)$, the source of field $\chi$ introduced in
the context of the large-N expansion. The generating functional
$Z[j]$ and connected Green's function generator $W[j]$ is given by
the path integral:
\begin{equation}
   Z[j]
   =
   e^{i \, W[j] }
   =
   \prod_{\alpha=0}^{N} \int \mathrm{d} \phi_\alpha \,
   e^{i \, S[\phi;j]}
   \label{SD.eq:Z}.
\end{equation}
The path integral needs to be supplemented by stating the boundary
conditions on the Green's functions. For initial-value problems
one needs  the CTP  boundary conditions~\cite{CTP} where in the
time integration the time contour is a closed time path contour.
However in discussing renormalization we only need to  consider
the vacuum equations and the Feynman boundary conditions on the
fields.  This is achieved by the usual $i \epsilon$ prescription
in deforming time slightly into the Euclidean region. The action
$S[\phi;j]$ is given by:
\begin{align}
   S[\phi;j]
   = &
   - \frac{1}{2} 
     \int \mathrm{d} x \!\! \int \mathrm{d} x' \,
   \phi_\alpha(x) \, \Delta_{\alpha \beta}^{-1}[\phi](x,x') \, \phi_\beta(x')
   \notag \\ &
   +
   \int \mathrm{d} x \, \phi_\alpha(x) \, j_\alpha(x) \>.
   \label{eq:SNxj}
\end{align}

Here, we have introduced the notation
\begin{equation}
   \Delta_{\alpha \beta}^{-1}[\phi](x,x')
   =
   \begin{pmatrix}
      D_0^{-1}(x,x') & 0                  \\
      0              & G_{0 \, ij}^{-1}(x,x')
   \end{pmatrix}  \>,
   \label{SD.eq:Ginvdef}
\end{equation}
with
\begin{align}
   D_0^{-1}(x,x')
   &=
      - \, \frac{1}{g} \ \delta(x,x') \>,
   \notag \\
   G_{0 \, ij}^{-1}(x,x')
   &=
   \Bigl [
      \Box
      + \chi(x)
   \Bigr ] \, \delta_{ij} \delta(x,x')
   \>.
   \label{SD.eq:dginv}
\end{align}
The above are the diagonal entries in the Green's function matrix
$G^{-1}_{0 \, \alpha \beta}[\phi](x,x')$ defined as follows:
\begin{align}
   G_{0 \, \alpha \beta}^{-1}[\phi](x,x')
   = &
   - \frac{ \delta^2 S[\phi;j] }
          {\delta \phi_{\alpha}(x) \, \delta \phi_{\beta}(x') }
\label{eq:g0invdef}
   \\ \notag
   = &
   \begin{pmatrix}
      D_0^{-1}(x,x')   & \bar{K}_{0 \, j}^{-1}(x,x') \\
      K_{0 \, i}^{-1}(x,x') & G_{0 \, i j}^{-1}(x,x')
   \end{pmatrix}  \>,
\end{align}
while the off-diagonal elements are
\begin{equation}
   K_{0 \,i}^{-1}[\phi](x,x') \ = \
   \bar{K}_{0 \, i}^{-1}[\phi](x,x') \ = \
   \phi_i(x) \, \delta(x,x')
   \>.
\end{equation}

The S-D equations are obtained from the identity
\begin{equation}
   \prod_{\beta=0}^{N} \int \mathrm{d} \phi_\beta \
   \frac{\delta}{\delta \phi_\alpha(x)} \
   e^{i \, S[\phi;j]}
   \ = \
   0
   \>.
\end{equation}
The Heisenberg equations of motion and constraint describing the
time evolution of the O(N) model are obtained as
\begin{gather}
   - \frac{1}{g} \, \chi(x)
   +
   \frac{1}{2}
      \sum_i \
      \Bigl [
         \phi_i^2(x)
         +
         G_{ii}(x,x)/i
      \Bigr ]
      -
   \frac{\mu^2}{g}
   =
   J(x)  \>,
   \label{SD.eq:chieq}
   \\
   \Bigl [
      \Box
      + \chi(x)
   \Bigr ] \, \phi_i(x)
   +
   K_i(x,x) / i
   =
   j_i(x) \>.
   \label{SD.eq:phieq}
\end{gather}
where the Green's functions $G_{\alpha \beta}[j](x,x')$ are
defined by:
\begin{align}
   G_{\alpha \beta}[j](x,x')
   = &
   \frac{\delta \phi_{\alpha}(x)}{\delta j_{\beta}(x')}
   =
   \frac{\delta^2 W[j]}{\delta j_\alpha(x) \, \delta j_\beta(x')}
   \label{SD.eq:GGdef}
   \\ \notag
   = &
   \begin{pmatrix}
      D(x,x')     & K_j(x,x') \\
      \bar K_i(x,x') & G_{i j}(x,x')
   \end{pmatrix}  \>.
\end{align}
Next, we introduce the 1-PI generating functional or effective
action $\Gamma[\phi]$ by performing the Legendre transformation
\begin{equation}
   \Gamma[\phi]
   =
   W[j]
   -
   \int \mathrm{d} x \,  \phi_\alpha(x) j_\alpha(x)
   \>.
\end{equation}
We obtain the equations of motion and constraint
\begin{align}
   - \frac{\delta \Gamma[\phi]}{\delta \chi(x)}
   & = J(x)
   \\ \notag
   & =
   - \frac{1}{g} \, \chi(x)
   +
   \frac{1}{2}
      \sum_i
   \Bigl [
         \phi_i^2(x)
         +
         G_{ii}(x,x)/i
   \Bigr ]
      -
      \frac{\mu^2}{g}
   \>,
   \\
   - \frac{\delta \Gamma[\phi]}{\delta \phi_i(x)}
   & = j_i(x)
   \\ \notag
   & =
   \Bigl [
      \Box
      + \chi(x)
   \Bigr ] \, \phi_i(x)
   +
   K_i(x,x) / i\>.
\end{align}
We also define the inverse Green's functions
\begin{align}
   G_{\alpha \beta}^{-1}[\phi](x,x')
   = &
   \frac{\delta j_\alpha(x)}{\delta \phi_\beta(x')}
   =
   - \frac{\delta^2 \Gamma[\phi]}{\delta \phi_\alpha(x) \, \delta \phi_\beta(x')}
   \\ \notag
   = &
   \begin{pmatrix}
      D^{-1}(x,x')   & \bar{K}_j^{-1}(x,x') \\
      K_i^{-1}(x,x') & G_{i j}^{-1}(x,x')
   \end{pmatrix}  \>,
\end{align}
such that
\begin{equation}
   \int \mathrm{d} x''
   G_{\alpha \beta}^{-1}[\phi](x,x'') \,
   G_{\beta \gamma}[j](x'',x')
   =
   \delta_{\alpha \gamma} \delta(x,x')
\label{eq:invert}
   \>.
\end{equation}
The Green's functions $G_{\alpha \beta}$ are obtained by inverting
the equation
\begin{equation}
   G_{\alpha \beta}^{-1}(x,x')
   =
   G_{0 \, \alpha \beta}^{-1}(x,x')
   +
   \Sigma_{\alpha \beta}(x,x')  \>,
   \label{eq:GGinvGinvSigma}
\end{equation}
where $G_{0 \, \alpha \beta}^{-1}(x,x')$ is given by
Eq.~\eqref{eq:g0invdef}, and we have introduced the generalized
self-energy matrix $\Sigma_{\alpha \beta}(x,x')$ as
\begin{equation}
   \Sigma_{\alpha \beta}[\phi](x,x')
   =
   \begin{pmatrix}
      \Pi(x,x')          & \Omega_j(x,x') \\
      \bar\Omega_i(x,x') & \Sigma_{ij}(x,x')
   \end{pmatrix}  \>.
   \label{eq:Sigmasdefs}
\end{equation}
By definition, the elements of the self-energy matrix are given as
\begin{align}
   \Sigma_{00} \ \rightarrow \
   \Pi(x,x')
   &=
      \frac{1}{2i} \,
   \frac{ \delta G_{jj}(x',x') }
        { \delta \chi(x) }
   \>,
   \\
   \Sigma_{i0} \ \rightarrow \
   \bar \Omega_i(x,x')
   &=
      \frac{1}{2i} \,
   \frac{ \delta G_{jj}(x',x') }
        { \delta \phi_i(x) }
   \>,
   \notag \\
   \Sigma_{0j} \ \rightarrow \
   \Omega_j(x,x')
   &=
      \frac{1}{i} \,
   \frac{ \delta K_j(x',x') }
        { \delta \chi(x) }
   \>,
   \notag \\
   \Sigma_{ij} \ \rightarrow \
   \Sigma_{ij}(x,x')
   &=
      \frac{1}{i} \,
   \frac{ \delta K_j(x',x') }
        { \delta \phi_i(x) }
   \>.
   \notag
\end{align}
The elements of the self-energy matrix are obtained by taking the
functional derivative of Eq.~\eqref{eq:invert}. We have
\begin{align}
   \frac{\delta G_{\alpha \beta}[j](x_1,x_2)}
        {\delta \phi_\gamma(x_3)}
   = &
   -
   \int \mathrm{d} x_4 \int \mathrm{d} x_5
   G_{\alpha \delta}[j](x_1,x_4)
\label{e:derivG}
   \\ \notag & \times
   \Gamma_{\delta \epsilon \gamma}[\phi](x_4,x_5,x_3)
   G_{\epsilon \beta}[j](x_5,x_2) \>.
\end{align}
Here, $\Gamma_{\alpha\beta\gamma}$ denotes the three-point vertex
function
\begin{align}
   \Gamma_{\alpha \beta \gamma}[\phi](x_1,x_2,x_3)
   = &
   \frac{\delta G_{\alpha \beta}^{-1}[\phi](x_1,x_2)}
        {\delta \phi_\gamma(x_3)}
   \\ \notag
   = &
   - \frac{\delta^3 \Gamma[\phi]}
          {\delta \phi_\alpha(x_1) \, \delta \phi_\beta(x_2) \,
           \delta \phi_\gamma(x_3)} \>.
\end{align}

In general, the three-point vertex function is the solution of the
exact equation
\begin{equation}
   \Gamma_{\alpha\beta\gamma}(x,x',x'')
   =
   f_{\alpha\beta\gamma} \, \delta(x,x') \, \delta(x,x'')
   +
   \frac{ \delta \, \Sigma_{\alpha\beta}(x,x') }
        { \delta \phi_{\gamma}(x'') } \>,
   \label{e:vert}
\end{equation}
where $f_{0,i,j} = f_{i,0,j} = f_{i,j,0} = \delta_{ij}$. The
self-energy matrix depends implicitly on the three-point function
as shown in Eq.~\eqref{e:derivG}. From this equation, one can
write a S-D equation for $\Gamma$ in terms of an irreducible
scattering kernel.  We will derive this below in the unbroken
symmetry case.

%
%
\subsection{Unbroken Symmetry Case}

These equations simplify considerably for the \emph{symmetric}
case, i.e. when $\phi(x) = 0$. The Green's function matrix is
diagonal, $G_{ii}(x,x') = G(x,x') \, \delta_{ij}$. The self-energy
matrix is also diagonal, $\Omega_i(x,x') = \bar \Omega_i(x,x') =
0$, and $\Sigma_{ij}(x,x') = \Sigma(x,x') \, \delta_{ij}$. We have
the gap equation
\begin{gather}
   \chi(x)
   =
   - \,
   \mu^2
   \ + \
   \frac{\lambda}{2i} \ G(x,x)
   \>,
   \label{eq:gap}
\end{gather}
and the Green's functions:
\begin{align}
   D^{-1}(x,x') \ = \ &
   - \, \frac{N}{\lambda} \, \delta(x,x')
   \ + \ \Pi(x,x')
   \>,
   \\
   G^{-1}(x,x') \ = \ &
   \Bigl [
      \Box
      + \chi(x)
   \Bigr ] \, \delta(x,x')
   \ + \ \Sigma(x,x')
   \>,
\end{align}
where, by definition, the polarization and self-energy are
\begin{align}
   \Pi(x,x')
   &=
   \frac{iN}{2}
   \int \!\! \mathrm{d} x_1 \!\! \int \!\! \mathrm{d} x_2 \,
   G(x,x_1)
   \Gamma(x_1,x_2,x'')
   G(x_2,x')
   \>,
   \\
   \Sigma(x,x')
   &=
   i
   \int \mathrm{d} x_1 \!\!\! \int \mathrm{d} x_2
   D(x,x_1)
   \Gamma(x_1,x_2,x'')
   G(x_2,x')
   \>.
\end{align}
We notice that $\Sigma$ is of order 1/N since $D$ is of order $1/N$.

In the symmetric case, there are S-D equations for the $\chi\phi
\phi$ vertex $\Gamma$  which needs renormalization: Functionally
differentiating the $\phi$ inverse propagator with respect to
$\chi(z)$  we can write
\begin{align}
&
   \Gamma(x,y,z) \ = \ \delta(x-y) \ \delta (x-z)
   \\ \notag &
   -i \int \mathrm{d}x_1 \mathrm{d}x_2 \mathrm{d}x_3  G(x-x_2) D(x-x_1) M(x_1,x_2,x_3,z)
   \>.
\end{align}
where $M$ is the  $\phi -\chi$ 1-PI scattering amplitude in the
\emph{s} channel $(x_1,x_2)$.  In this paper we will use the
schematic form
\begin{equation}
   \Gamma \ = \ 1 \ -\ i \, D M G
   \>,
\end{equation}
as shorthand for the above equation in either coordinate or
momentum space when appropriate.

The three graphs contributing to this are:
\begin{align}
   M(x_1,& x_2,x_3,x_4) \ = \ M_{stu}(x_1,x_2,x_3,x_4)
\label{eq:msut}
   \\ \notag &
   + \int \mathrm{d}x_5 \mathrm{d}x_6 \Gamma(x_2,x_3,x_5) D(x_5,x_6) \Lambda (x_6,x_1, x_4)
   \\ \notag &
   + \int \mathrm{d}x_5 \mathrm{d}x_6 \Gamma(x_1, x_4,x_5) G(x_5,x_6) \Gamma (x_5,x_2, x_3)
   \>,
\end{align}
where
\begin{equation}
   \Lambda(x_1,x_2,x_3) = \frac{\delta D^{-1}(x_1,x_2) }{\delta \chi(x_3)}
\end{equation}
is the 1-PI 3-$\chi$ vertex function which is  finite, and
\begin{equation}
   M_{stu}(x_1,x_2,x_3,x_4)= \frac{\delta \Gamma(x_1,x_2, x_3) } {\delta \chi(x_4)}
\end{equation}
$M_{stu}$ is  the 1-PI  in the  \emph{s}, \emph{t}, and \emph{u}
channels scattering amplitude for $\phi-\chi$ elastic scattering.
For renormalization purposes it is useful to have another
representation of $\Gamma$ in terms of $K$  the $2-PI$ in the
\emph{s} channel scattering kernel since this will facilitate the
renormalization program needed below.

%
%
\subsection{BVA}

In the bare-vertex approximation (BVA)~\cite{ref:noneq1}, the
three-point vertex function $\Gamma_{\alpha\beta\gamma}$  that
appears in the S-D equations for the inverse Green's function is
approximated by keeping only the contact term, i.e.
\begin{equation}
   \Gamma_{\alpha\beta\gamma}^{\text{(BVA)}}(x,x',x'')
   =
   f_{\alpha\beta\gamma} \, \delta(x,x') \, \delta(x,x'')
   \>.
\end{equation}
This is justified at large-N where the vertex corrections to the
inverse Green's function equations first appear at order~1/N$^2$.

In this approximation, we have
\begin{equation}
   \left [
   \frac{ \delta G_{\mu \nu}(x_1,x_2) }
        { \delta \phi_\gamma(x) }
   \right ]^{\text{(BVA)}}
   = - \ G_{\mu \alpha}(x_1,x) f_{\alpha \beta \gamma}
   G_{\beta \nu}(x,x_2)
   \>,
\end{equation}
which leads to the self-energies $\Sigma_{\alpha
\beta}^{\text{BVA}}$ given by
\begin{align}
   \Sigma_{00} \ \rightarrow \
   \Pi^{\text{(BVA)}}(x,x')
   &=
      \frac{i}{2} \
      G_{m n}(x,x') \ G_{m n}(x,x')
   \>,
   \label{e:SigmasBVA}
   \\
   \Sigma_{i0} \ \rightarrow \
   \bar \Omega_i^{\text{(BVA)}}(x,x')
   &=
      i \,
   G_{i m}(x,x') \ K_m(x,x')
   \>,
   \notag
   \\
   \Sigma_{0j} \ \rightarrow \
   \Omega_j^{\text{(BVA)}}(x,x')
   &=
      i \,
      \bar{K}_m(x,x') \ G_{m j}(x,x')
   \>,
   \notag
   \\
   \Sigma_{ij} \ \rightarrow \
   \Sigma_{ij}^{\text{(BVA)}}(x,x')
   &=
      i \,
   \Bigl [
   G_{i j}(x,x') \ D(x,x')
   \notag
   \\ & \qquad \quad
   +
   \bar{K}_i(x,x') \ K_j(x,x')
   \Bigr ]
   \>,
   \notag
\end{align}
where we have used the symmetry property, $G_{ij}(x,x') =
G_{ji}(x',x)$ and $K_i(x,x') = \bar K_i(x',x)$. It can be verified
by direct calculation, that indeed $\bar\Omega_i(x,x') =
\Omega_i(x',x)$, as expected.

To summarize, in the BVA, one solves the equations of motion for
$\phi_i(x)$
\begin{gather}
   \Bigl [
      \Box + \chi(x)
   \Bigr ] \, \phi_i(x)
   +
   K_i(x,x) / i
   = 0 \>,
   \label{eq:GammaxeqBVA}
\end{gather}
and the gap equation for $\chi(x)$
\begin{gather}
   \chi(x)
   =
   - \,
   \mu^2
   \ + \
   \frac{g}{2}
      \sum_i \
      \Bigl [
         \phi_i^2(x)
         +
         G_{ii}(x,x)/i
      \Bigr ]
   \>,
   \label{eq:GammachieqBVA}
\end{gather}
self-consistently with the equations for the Green's functions
\begin{equation}
   G_{\alpha \beta}^{-1}(x,x')
   =
   G_{0 \, \alpha \beta}^{-1}(x,x')
   +
   \Sigma_{\alpha \beta}^{\text{(BVA)}}(x,x')
   \>.
   \label{eq:GGinvGinvSigmaBVA}
\end{equation}

In the symmetric case the BVA polarization and self-energy are
simply
\begin{align}
   \Pi^{\text{(BVA)}}(x,x')
   &= \frac{iN}{2} \, G(x,x') \, G(x,x')
   \>,
   \label{eq:pi_0}
   \\
   \Sigma^{\text{(BVA)}}(x,x')
   &= i \, G(x,x') \, D(x,x')
   \label{eq:sig_0}
   \>.
\end{align}

%
%
\subsection{2-PI expansion}

Here we review the 2-PI generating functional~\cite{ref:2PI}.  We
then derive the S-D equations that follow when we include in
$\Gamma_2$ the first term in a symmetrical (in propagators $\chi
\phi$) loop expansion of the generator $\Gamma_2$  of the  2-PI
graphs, namely the two-loop graph.  To obtain a~1/N reexpansion,
one then recognizes that in this loop expansion, the three types
of propagators have different dependence on~1/N ($G \sim
\mathcal{O}(1)$ and $D, K \sim \mathcal{O}(1/N)$). The effective
action is the twice-Legendre transformed generating functional:
\begin{align}
   \Gamma[\phi, G]
   \ = \ &
   S_{\text{class}}[\phi] +
   \frac{i}{2} \mathrm{Tr} \{ \, \ln \, [ \, G^{-1} \, ] \}
   \label{eq:CJT}
   \\ \notag &
   +
   \frac{i}{2} \mathrm{Tr} \{ G_0^{-1}[\phi] \, G - 1 \}
   +
   \Gamma_2[G]  \>.
\end{align}
where
\begin{equation}
   G^{-1}_{0~\alpha\beta}[\phi](x,x')
   =
   - \frac{\delta^2 S_{\text{cl}}[\phi]}
          {\delta \phi_{\alpha}(x) \, \delta \phi_{\beta}(x')}
\end{equation}
For initial-value problems, $\phi$  is also a matrix in CTP space.
$S_{\text{class}}$ is the classical Lagrangian written in terms of
both $\phi$ and $\chi$. In the auxiliary field formalism the
propagators for $\phi \phi$,  $\phi \chi$ and $\chi \chi$ are
treated on the same footing. Thus if we make a loop expansion the
lowest order term in $\Gamma_2$ has two loops. In contrast, as
discussed before, the~1/N approximation is asymmetric in $\chi$
and $\phi$, since it is all order in $\phi$ (for fixed $ \chi$ and
a loop expansion only in $\chi$.

The equations of motion for the field expectation values follow by
variation of $\Gamma$ with respect to $\phi_i $ and $\chi$. We
find
\begin{equation}
 -\left[ \square_x+ \chi(x)\right] \phi_i(x) = { K}_i(x,x),
\end{equation}
and
\begin{equation}
\chi(x) = - \mu^2 + \frac{g}{2} \sum_i [\phi_i^2(x) + G_{ii}(x,x)].
\end{equation}
The equation for the two-point function follows by variation with
respect to ${G}$, which gives
\begin{equation}
   {G}^{-1}_{\alpha \beta} =  {G}^{-1}_{0\, \alpha \beta} + \Sigma_{\alpha \beta}
   \>,
\end{equation}
where
\begin{equation}
\label{sigma2}
   \Sigma_{\alpha \beta} \ = \ - \ 2i \ \frac{\delta \Gamma_2[{G}]}{\delta {G}_{\alpha \beta}}
   \>.
\end{equation}
At the two-loop level we have that
\begin{align}
   \Gamma_2[G] \ = \ & - \frac{1}{12} {\rm Tr} \{f GGGf \}
   \\ \notag &
   \ = \ - \frac{1}{4} \left[ G_{ij} G_{ij} D + 2 \bar K_i K_j G_{ij} \right]
   \>.
\end{align}
Taking the derivatives of $\Gamma_2$ with respect to $G_{\alpha
\beta}$, the $\Sigma$ matrix given by \eqref{sigma2} leads to the
same form for the matrix elements as we obtained previously from
the S-D equations in the BVA. We notice, since $D \sim$~1/N and $K
\sim$~1/N, the two contributions to $\Gamma_2$ are of order 1
and~$1/N$, respectively, whereas the classical action is of order
$N$. Thus if one is interested in counting powers of 1/N rather
than loops, one would in next-to-leading order in $1/N$ ignore the
second contribution to $\Gamma_2$.  This fact distinguishes
counting loops in the auxiliary field formalism  from counting
powers of $1/N$ as noted in Refs.~\cite{ref:noneq2, berges-2PI}.

The quantity $\Gamma_2[G]$ has a simple graphical interpretation
in terms of all the 2-PI vacuum graphs using vertices from the
interaction term $ -\frac{1}{2} \chi \phi_i \phi_i$.  When
$\langle \phi \rangle_0 = 0$, one obtains
\begin{align}
   \Gamma[\chi,G_\phi,D] &
   \ = \
   S_{\text{class}}[\chi]
   \label{eq:GammaDDSA}
   \\ \notag &
   +
   \frac{i}{2} \, \mathrm{Tr} \{ \, \ln \, [ \, D^{-1} \, ] \}
   +
   \frac{i}{2} \, \mathrm{Tr} \{ \, \ln \, [ \, G_\phi^{-1} \, ] \}
   \\ \notag &
   +
   \frac{i}{2} \, \mathrm{Tr}
      \{ D_0^{-1} \, D + G_0^{-1} \, G_\phi - 2 \}
   +
   \Gamma_2[G_\phi,D] \>.
\end{align}
where $G \equiv \{ G_\phi,D \}$ and
\begin{align}
&
   \Gamma_2[G_\phi,D]
   =
 \\ \notag &   - \frac{1}{4}
    \int \mathrm{d} x_1 \int \mathrm{d} x_2 \,
      D(x_1,x_2) \, G_\phi(x_1,x_2) \, G_\phi (x_2,x_1)  \>.
\end{align}
The resulting equations for the two-point functions have no vertex
corrections.


%
%
\section{All orders Renormalization in the Vacuum Sector}
\label{sec:allorders}

For renormalization it is necessary to only study the theory with
unbroken symmetry since the renormalization is not changed when
$\phi \neq 0$ (see e.g. Ref.~\cite{ref:renorm}).  The advantage of
studying $\phi^4$ theory in terms of the auxiliary field $\chi$ is
that the~1/N resummation improves the renormalizability as first
discussed by Gross~\cite{ref:Gross}. The renormalized theory can
then be determined in terms of two "physical" parameters, which we
will choose to be the value of the $\phi$  mass in the unbroken
vacuum and  value of the coupling constant at $q^2=0$. What we
will find is that only the $\phi$ two-point function needs wave
function renormalization and there is a Ward-like identity
relating this renormalization constant to the $\chi \phi \phi$
vertex renormalization constant. The important effect of
reexpressing $\phi^4$ field theory in terms of the $\chi$
propagator $D$  is that once the above renormalizations are
performed, there are no further divergences in the elastic
scattering of two $\phi$ particles.

%
%
\subsection{Momentum space representation}

Here we consider the homogeneous case, relevant to understanding
the vacuum sector.  In Minkowski space, we define Fourier
transforms as:
\begin{align}
   G(x,x')
   =
   \int [ \mathrm{d} p ] \, e^{- i p \cdot (x - x')} \ G(p)
   \>,
   \\
   D(x,x')
   =
   \int [ \mathrm{d} p ] \, e^{- i p \cdot (x - x')} \ D(p)
   \>,
\end{align}
where $[ \mathrm{d} p ] = \frac{ \mathrm{d}^{d} p }{ ( 2\pi
)^{d}}$. Then, the gap equation is written as
\begin{align}
   \chi
   &=
   - \,
   \mu^2
   \ + \
   \frac{g}{2i} \ \int [ \mathrm{d} p ] \, G(p)
   \>,
\end{align}
and the equations for the Green's functions as
\begin{align}
   D^{-1}(p) \ = \ &
   - \, \frac{N}{\lambda}
   \ + \ \Pi(p)
   \>,
   \\
   G^{-1}(p) \ = \ &
   \bigl [
      - p^2 + \chi
   \bigr ]
   \ + \ \Sigma(p)
   \>.
\end{align}
Finally, the polarization and self-energy are given by
\begin{align}
   \Pi(p)
   & =
  \frac{i N}{2} \int [ \mathrm{d} q ] \, G(q) \, \Gamma(q,p-q) \, G(p-q)
   \>,
   \\
   \Sigma(p)
   & =
   {i} \int [ \mathrm{d} q ] \, D(q) \, \Gamma(q,p-q) \, G(p-q)
   \>.
\end{align}

The only primitive vertex function for this theory is the $\chi
\phi \phi$ vertex which we will just call $\Gamma= \Gamma_{ii0}$.
Keeping  in mind the translational invariance properties, we write
\begin{align}
   \Gamma(x,x',x'')
   &=
   \int [ \mathrm{d} p ]
   e^{ -i p \cdot ( x - x'')}
   \!\!
   \int [ \mathrm{d} q ]
   e^{ -i q \cdot ( x' - x'' )}
   \Gamma(p,q)
   \>,
\end{align}
and the vertex equation~\eqref{e:vert} becomes
\begin{align}
   \Gamma(p,q)
   \ = \ 1 \ + \
   \Delta \Gamma(p,q)
   \>.
\end{align}
The original S-D equation for $\Gamma$ can be rewritten in terms
of a 2-PI scattering kernel $K_{2}$, which we schematically write
in matrix form as
\begin{equation}
   \Gamma(p,p+q) = 1 \ + \ i [\Gamma  DK_2G] (p, p+q)
   \>.
\end{equation}

The primitive divergences of this theory have been discussed in
detail in Ref.~\cite{bcg}. The minimal degree of divergence
(ignoring $\log$ improvements) is given by the simple formula
\begin{equation}
   \mathrm{D} \ = \ 4 \ - \ 2\, \mathrm{B} \ - \ \mathrm{M}
   \>,
\end{equation}
where $\mathrm{B}$ is the number of external auxiliary field
propagators $D$, and $\mathrm{M}$ is the number of external meson
propagator ($G$) lines. Thus the meson propagator is naively
divergent as $\Lambda^2$ and needs two subtractions (mass and wave
function renormalization), the $D$ propagator is $\log$ divergent
and needs one subtraction (coupling constant renormalization), and
the vertex function has D$=0$ and is $\log\ \log$ divergent.
Another potentially divergent graph is the graph having four
external meson lines, which in principal could be $\log$
divergent. However, the $D$ propagators actually go as
$1/\ln(p^2)$ at high momentum, and so the box graph with two D and
two G propagators actually converges since
\begin{equation}
   \int^\Lambda \frac {\mathrm{d}x}{x (\ln x)^2}  \sim \frac{1} {ln \Lambda}
   \>.
\end{equation}

In this paper all the renormalizations will take place on mass
shell, and we follow the treatment in~\cite{CHKMPA}, since that
approach is easiest to implement in the time evolution problem,
with the power series in $-q^2+m^2$ becoming $\Box + m^2$ acting
on lattice versions of $\delta$ functions. Our renormalization
procedure will involve two steps.  First we will identify the wave
function and vertex renormalization constants as well as the
physical mass of the $\phi$ particle. We will then obtain naively
finite equations for the multiplicatively renormalized propagators
and vertex functions.  Secondly we will  show that when we replace
the bare vertices by the full vertex function minus a correction,
we obtain finite renormalization constant free equations for the
renormalized Green's functions and vertex functions.

Before proceeding, let us define the multiplicative
renormalization constants.  We first identify the physical mass by
the zero of the inverse propagator for the $\phi$ field,
\begin{equation}
   G^{-1}(p^2 = m^2) \ = \ 0
   \>.
\end{equation}
Once this mass is identified, then the wave function
renormalization constant for $\phi$ is defined as
\begin{equation}
   Z_2^{-1}(m^2) \ = \ - \ \frac{\mathrm{d}G^{-1}(p^2)} {\mathrm{d}p^2} \Bigr |_{p^2 = m^2}
   \label{eq:zzz2}
   \>.
\end{equation}
If we write $G^{-1} = -p^2 + \chi +\Sigma$, then  we also can
write Eq.~\eqref{eq:zzz2} as
\begin{equation}
   Z_2^{-1}(m^2) \ \equiv \
   1 \ - \ \frac{\mathrm{d}\Sigma(p^2)} {\mathrm{d}p^2} \Bigr |_{p^2 = m^2}
   \>.
\end{equation}
The renormalized $\phi$ propagator is then defined by
\begin{equation}
   G_R(p^2) \ = \ Z_2^{-1} \ G(p^2)
   \>.
\end{equation}

The vertex function renormalization constant is defined by
\begin{equation}
   \Gamma_R(p, p+q) \ = \ Z_1 \ \Gamma(p, p+q)
   \>,
\end{equation}
with the condition that on mass shell, with no momentum transfer.
We have
\begin{equation}
   \Gamma_R(p,p)|_{p^2=m^2} \ = \ 1
   \>,
\end{equation}
which leads to
\begin{equation}
   Z_1^{-1} \ = \
   \Gamma(p,p) |_{p^2=m^2} \ \equiv \
   1 \ + \ \frac{\partial \Sigma(p^2)}{\partial \chi} \Bigr |_{p^2=m^2}
   \>.
\end{equation}
We will prove later that a Ward-like identity leads to the
renormalization constants being equal, i.e. $Z_1= Z_2$, which
tells us that the product $\Gamma G$ is renormalization scheme
invariant.

%
%
\subsection{Analyzing Divergences}

First let us realize that with our way of writing the Lagrangian,
$D(q^2)$ is a renormalization group (RG) invariant and is just
(apart from a sign)  the  running renormalized coupling constant
$g_r(q^2)$ for scalar meson scattering via  single $\chi$ meson
exchange. Thus $g_r(0) \equiv g_r$ is related to the scattering at
$q^2=0$. In weak coupling, or in leading order in large-N, $g_r$
is the actual value of the scattering amplitude. However, in the
full theory the scattering amplitude gets corrections from all the
loops involving exchanging more and more $D$ propagators. We have
\begin{equation}
   D^{-1}(q^2=0 ) \ = \
   - \, \frac{1}{g_r}
   \ = \
   - \, \frac{1}{g} \ + \ \Pi(q^2=0)
   \>,
\end{equation}
so that
\begin{equation}
   - \, \frac{1}{g} \ = \ - \, \frac{1}{g_r} \ - \ \Pi(q^2=0)
   \label{eq:gtogr}
   \>.
\end{equation}
If we wish to define a coupling constant renormalization constant via
\begin{equation}
g= Z_g^{-1} g_R
\end{equation}
then
\begin{equation}
Z_g = 1+ g_R \Pi(0)
\end{equation}
In terms of $g_R$, the expression for
the inverse propagator is
\begin{equation}
   D^{-1}(q^2) = - \frac{1}{g_R} + \Pi^{\mathrm{[sub\, 1]}}(q^2)
   \>,
\end{equation}
with
\begin{equation}
   \Pi^{\mathrm{[sub\, 1]}}(q^2) \ = \ \Pi(q^2) - \Pi(0)
   \>.
\end{equation}
Since $\Pi(q^2)$ is naively $\log$ divergent, then
$\Pi^{\mathrm{[sub\, 1]}}(q^2)$ is naively finite. At this stage
we have that $\Pi = G\Gamma G$, which is not yet in a form which
displays its independence of the renormalization constants.

Now let us look at the divergences of the $\phi$ propagator. If we
write $G^{-1} = -p^2 + \chi +\Sigma$, with $\Sigma = G \Gamma D$,
and
\begin{equation}
   \chi \ = \ -\ \mu^2 \ + \ \frac{\lambda}{2i} \ G(x,x)
   \>,
\end{equation}
then the bubble $G(x,x)$ has quadratic as well as $\log$
divergences in $3+1$, which are related to the mass and coupling
constant renormalizations, respectively. In turn, the self-energy
$\Sigma$ has quadratic and $\log\ \log$ divergences related to
mass and wave function renormalization. Identifying the physical
mass yields the relationship
\begin{equation}
   0 \ = \ - \, m^2 \ + \ \chi \ + \ \Sigma(p^2=m^2)
   \>.
\end{equation}
To identify the naively  finite part of $\Sigma$ we now expand
$\Sigma$ around $p^2 = m^2$
\begin{equation}
   \Sigma(p^2) \ \equiv \ \Sigma_0 \ + \
   \Sigma_1 (p^2-m^2) \ + \ \Sigma^{\mathrm{[sub\, 2]}}(p^2)
   \>,
\end{equation}
where  $\Sigma_0 =\Sigma(p^2=m^2)$,  $\Sigma_1 =
\frac{\mathrm{d}\Sigma}{\mathrm{d}p^2} |_{(p^2=m^2)}$, and
$\Sigma^{\mathrm{[sub\, 2]}}(p^2)$ is naively finite and vanishes
quadratically at the physical mass. Identifying
\begin{equation}
   Z_2^{-1}(m^2) \ \equiv \ - \frac{\mathrm{d} G^{-1}(p^2)}{\mathrm{d} p^2} \Bigr |_{p^2=m^2}
   \>,
\end{equation}
we can write
\begin{equation}
   G^{-1} \ = \ Z_2^{-1} \left( - \, p^2 \ + \ m^2 \ + \ Z_2 \Sigma^{\mathrm{[sub\, 2]}} \right)
   \>,
\end{equation}
so that
\begin{equation}
   G_R^{-1} \ = \ - \, p^2 \ + \ m^2 \ + \ Z_2 \Sigma^{\mathrm{[sub\, 2]}}
   \>.
\end{equation}
This equation is now naively finite but not transparently
independent of the renormalization constants.  We will have to
symmetrize the Dyson equation for $\Sigma$ with respect to having
fully dressed vertices at both ends in order to do this.

Next we study the divergence of the vertex function. The vertex
function can be written as
\begin{equation}
   \Gamma(p,q) \ = \ \Gamma(p,p)|_{p^2=m^2} \ + \ \Gamma^{\mathrm{[sub\, 1]}}(p,q)
   \>,
\end{equation}
where $\Gamma^{\mathrm{[sub\, 1]}}(p,q) = \Gamma(p,q) -
\Gamma(p,p)|_{p^2=m^2}$ is the vertex function once subtracted on
the mass shell. Naively, the first term is $\log\ \log$ divergent,
and so $\Gamma^{\mathrm{[sub\, 1]}}(p,q)$ is naively finite. The
renormalized vertex function on the mass shell with no momentum
transfer is defined to be one, giving the equation
\begin{equation}
   \Gamma_R(p,p) |_{p^2=m^2}
   \ = \ Z_1 \ \Gamma(p,p) |_{p^2=m^2}
   \ = \ 1
   \>,
\end{equation}
or
\begin{equation}
   Z_1^{-1} \ = \ \Gamma(p,p) |_{p^2=m^2}
   \>.
\end{equation}
 From this we obtain
\begin{equation}
   \Gamma_R(p,q) \ = \
   Z_1 \ \Gamma(p,q)
   \ = \
   1 \ + \ Z_1 \ \Gamma^{\mathrm{[sub\, 1]}}(p,q)
   \>.
\end{equation}
If we write the S-D equation for $\Gamma$ in the form $\Gamma = 1
+ \Delta \Gamma$ then we can rewrite this as
\begin{equation}
   \Gamma_R(p,q) \ = \ 1 \ + \ Z_1 \ \Delta \Gamma^{\mathrm{[sub\, 1]}}(p,q)
   \>,
\end{equation}
where the superscript means once subtracted on the mass shell.
Again we need to show that the right hand side, when symmetrized
appropriately, is independent of all the renormalization
constants.

%
%
\subsection{Obtaining finite equations for the renormalized Green's Functions}

To remove the dependence of the above equation on $Z_1$ and $Z_2$
the key ingredient is the symmetrization of the S-D equations,
which usually have one bare and one full vertex function in their
definition. We also need to prove the Ward-like  identity that
$Z_1=Z_2$, which is done as follows: We have already said that $D$
is RG invariant, so that the vacuum  vertex function satisfies
(for constant field $\chi$)
\begin{align}
   \Gamma(p,p)  \ = \ &
   \frac {\delta G^{-1}(p^2)}{\delta \chi}
   \ = \
   Z_2^{-1} \ \frac {\delta G_R^{-1}(p^2)}{\delta \chi}
   \notag \\
   \ \equiv \ &
   Z_1^{-1} \ \Gamma_R(p,p)
   \>.
\end{align}
However, since $\chi$ is not renormalized, we also have
\begin{equation}
   \Gamma_R(p,p) \ = \ \frac {\delta G_R^{-1}(p^2)}{\delta \chi}
   \>.
\end{equation}
Thus we find that in this theory we have the identity
\begin{equation}
   Z_1 \ = \ Z_2
   \>.
\end{equation}

This can also be shown by analyzing graphs. We have
\begin{align}
   Z_1^{-1}
   & \ = \
   \Gamma(p,p) |_{p^2=m^2}
   \ \equiv \
   1 \ + \ \frac {\partial \Sigma(p^2)}{\partial \chi} \Bigr |_{p^2=m^2}
   \>,
   \\
   Z_2^{-1}
   & \ = \
   1 \ - \ \frac {\partial \Sigma(p^2)}{\partial p^2} \Bigr |_{p^2=m^2}
   \>.
\end{align}
By studying graphs contributing to $ Z_1$ and $Z_2$ the difference
between these graphs are seen to be naively finite~ \cite{bcg},
and since these renormalization constants diverge as $\log\ \log
\Lambda$ one has in the continuum $Z_1= Z_2$.   From this identity
we find that the quantity $\Gamma G$ is renormalization scheme
invariant and equals $\Gamma_R G_R$.  We also have, $Z_1 \Gamma =
Z_2 \Gamma = \Gamma_R$.

The next step is to get an integral equation for $\Gamma$ which is
iterative in $\Gamma$. The original integral equation for $\Gamma$
that one gets by functional differentiation of the equation for
the inverse $\chi$ propagator is schematically
\begin{equation}
   \Gamma \ = \ 1 \ - \ i \, D M G
   \label{eq:Gaminv}
   \>,
\end{equation}
where $M$ is the one 1-PI in \emph{stu} $\phi \chi$ scattering
amplitude defined in Eq.~\eqref{eq:msut}.  We want to replace this
equation by one in which the r.h.s also has a $\Gamma$.

First we invert Eq.~\eqref{eq:Gaminv} to obtain the formal
identity (see Ref.~\cite{bcg})
\begin{equation}
   1 \ = \ \Gamma \ (1 \ - \ i \, D M G)^{-1}
   \>.
\end{equation}
We can therefore introduce a $\Gamma$ into the integral equation
using this identity:
\begin{align}
   \Gamma \ = \ &
   1 \ + \ i \, \Gamma \ (1 - i D M G)^{-1} DMG
   \\ \notag
   \ \equiv \ &
   1 \ - \ i \, \Gamma \ D K_2 G
\end{align}
Solving for $K_2$ and reintroducing the coordinates, we have
\begin{align}
   &
   M(x_1,x_2,x_3,x_4)
   \ = \ K_2(x_1,x_2,x_3,x_4)
   \\ \notag &
   \ - \
   \int \ \mathrm{d}x_5 \mathrm{d}x_6 \mathrm{d}x_7 \mathrm{d}x_8 \
   M(x_1,x_2,x_5,x_6) D(x_5,x_7)
   \\ \notag & \qquad \qquad \times \
   G(x_6,x_8) K_2(x_7,x_8,x_3,x_4)
   \>,
\end{align}
so that $K_2$ is the 2-PI in the \emph{s} channel irreducible
kernel of the Bethe-Salpeter equation for the scattering. What is
important for renormalization is that the combination $G D K$ is
RG invariant as can be seen by looking at all the skeleton terms
contributing to $K$.  For example, if the skeleton one-particle
exchange in the \emph{t} channel contribution to $K$, $\Gamma G
\Gamma$, is investigated, we get the combination $D \Gamma G
\Gamma G$, which is obviously equal to $D_R  \Gamma_R G_R \Gamma
G_R$.

The key to eliminating the dependence of the renormalized Green's
function equations obtained earlier on $Z_1$ and $Z_2$  is the
symmetrizing of the self-energy and polarization graphs with
respect to the full vertex function. The strategy for doing this
in quantum electrodynamics (QED) is found in the text book of
Bjorken and Drell~\cite{bd2}, where it is also shown that this
procedure eliminates the problem of overlapping divergences.

We have shown we can write
\begin{equation}
   \Gamma \ = \ 1 \ - \ i \, \Gamma  D K_2 G
   \ = \ 1 \ + \ \Delta \Gamma
   \>.  \label{Gamma}
\end{equation}
This equation allows us to substitute the bare vertex by
\begin{equation}
   1 \ = \ \Gamma - i \Gamma  DK_2G
   \ \equiv \
   \Gamma \ - \ \Delta \Gamma
   \>.
\end{equation}
The right hand side now multiplicatively renormalizes exactly as
$\Gamma$ does, by the above argument. Replacing bare vertices by
the right hand side will exactly absorb the extra factors of $Z_1$
and $Z_2$. Equation~\eqref{Gamma} is useful to symmetrize $\Sigma$
with respect to $\Gamma$.  However a different S-D equation for
$\Gamma$ is needed when we symmetrize $\Pi$ which is a loop made
of two $\phi$ propagators. To obtain this S-D equation we realize
that we can write $D^{-1}$ as
\begin{equation}
D^{-1}(x_1,x_2) = [\Box + \chi(x_1)] \delta(x_1-x_2) + \Sigma[G,D](x_1,x_2)
\end{equation}
where
\begin{equation}
 \Sigma[G,D](x_1,x_2)= -2i \frac{\delta \Gamma_2[G,D]}{\delta G(x_1,x_2)}
\end{equation}
is just a function of the exact $G$ and $D$ propagators.  Thus,
using the chain rule and taking the total functional derivative of
$D^{-1}$ with respect to $\chi(x_3)$, we obtain the equation
\begin{align}
&
\Gamma(x_1,x_2,x_3) = \delta(x_1-x_3) \delta(x_1-x_2)
- \! \int \! \mathrm{d}x_3 \mathrm{d}x_4 \mathrm{d}x_5 \mathrm{d}x_6
\notag \\ &
\times
\Bigl [ \Gamma(x_3,x_5,x_6) G(x_5, x_3)  M_2(x_3,x_4, x_1,x_2) G(x_4,x_6)
\notag \\ & \
+ \Lambda_3(x_3,x_5,x_6) D(x_5,x_3)K_2(x_3,x_4,x_1,x_2) D(x_4,x_6)
\Bigr ] \>.
\label{Gamma2}
\end{align}
In the above we have  that $\Lambda_3 = {\delta D^{-1}}/{\delta
\chi}$ is the three $\chi$ 1-PI vertex function.  The two
scattering kernels are that  $M_2$ is the 2-PI scattering kernel
for $\phi \phi \rightarrow \phi \phi$ and $K_2$ here is now the
\emph{t} channel kernel for $\chi \chi \rightarrow \phi \phi$.
Explicitly, we have
\begin{eqnarray}
M_2(x_1,x_2,x_3,x_4)&& = \frac{\delta G^{-1}(x_1,x_2) }{\delta G(x_3,x_4) } \nonumber \\
K_2(x_1,x_2,x_3,x_4)&& = \frac{\delta G^{-1}(x_1,x_2) }{\delta
D(x_3,x_4)} \>.
\end{eqnarray}

We note that in QED, $\Lambda_3$  would be the three-photon vertex
which is zero by Furry's theorem~\cite{furry}.  Again a study of
graphs shows that $\Delta \Gamma$ is rendered independent of
renormalization constants when multiplied by $Z_1=Z_2$. First, let
us look at the Dyson equation for the renormalized vertex
function. We use
\begin{equation}
   \Gamma(p,q) \ = \ 1 \ + \ \Delta \Gamma(p,q)
   \>,
\end{equation}
where $\Delta \Gamma(p,q) = i \, [\Gamma G D K_2] (p,q)$.
Subtracting once on the mass shell leads to
\begin{align}
   \Gamma_R (p,q) \ = \ Z_1 \Gamma(p,q)
   \ = \ & [ 1+ Z_1  \Delta \Gamma^{\mathrm{[sub\, 1]}}](p,q)
   \notag \\
   \ \equiv \ &
   1+  \Delta \Gamma_R(p,q)
   \>,
\end{align}
which is now explicitly finite and free from any dependence on the
renormalization constants, since
\begin{equation}
   \Delta \Gamma_R(p,q)
   \ = \
   [ \Gamma_RG_R D_R K_{2R}]^{\mathrm{[sub\, 1]}}(p,q)
   \>,
\end{equation}
which is finite and written in terms of a renormalized skeleton
expansion for the scattering kernel $K_{2R}$.  Similarly using the
second S-D equation we obtain
\begin{align}
   \Delta \Gamma_R(p,q)
   \ = \ &
   [ \Gamma_RG_R G_R M_{2R}]^{\mathrm{[sub\, 1]}}(p,q)
   \\ \notag &
   +  [ \Lambda_{3R}D_R D_R K_{2R}]^{\mathrm{[sub\, 1]}}(p,q)
   \>,
\end{align}
since $D_R \sim \mathcal{O}(1/N)$ and $\Delta \Gamma_R(p,q) \sim
\mathcal{O}(1/N)$.

The equation for the $ \chi$ propagator contains the naively
finite once subtracted polarization $ \Pi^{\mathrm{[sub\, 1]}} =
\left[ G \Gamma G \right]^{\mathrm{[sub\, 1]}} $. The right hand
side of this equation however is not RG invariant. By inserting
the second S-D equation for $\Gamma$ in the form $1 = \Gamma -
\Delta \Gamma$, then we have
\begin{align}
   \Pi^{\mathrm{[sub\, 1]}}
   \ = \ &
   [(\Gamma - \Delta \Gamma) G  G \Gamma]^{\mathrm{[sub\, 1]}}
   \\ \notag
   \ \rightarrow \ &
   [(\Gamma_R - \Delta \Gamma_R) G_R G_R \Gamma_R ]^{\mathrm{[sub\, 1]}}
   \>,
\end{align}
and we see  that all the dependence on the renormalization
constants disappears.

Next let us look at the $\phi$ propagator. We have previously
shown that
\begin{equation}
   G_R^{-1} (p^2) \ = \ m^2 - p^2 \ + \ Z_2 \Sigma^{\mathrm{[sub\, 2]}}(p^2)
   \>,
\end{equation}
where $\Sigma^{\mathrm{[sub\, 2]}}(p^2)$ is the twice subtracted
$G \Gamma D$ and so is naively finite. So it is easy to see that
once we replace the bare vertex by  the first S-D equation for $
\Gamma$ in the form $1 = \Gamma -  \Delta \Gamma$ in $\Sigma$ we
again have
\begin{equation}
   G^{-1} _R (q^2) \ = \ m^2 - p^2 \ + \ \Sigma_R^{\mathrm{[sub\, 2]}}(p^2)
   \>,
\end{equation}
with
\begin{align}
   \Sigma_R^{\mathrm{[sub\, 2]}}
   \ = \ &
   Z_2 \ \Sigma^{\mathrm{[sub\, 2]}}
   \\ \notag
   \ = \ &
   [(\Gamma_R -   \Delta \Gamma_R )G_R  D_R \Gamma_R]^{\mathrm{[sub\, 2]}}
   \>,
\end{align}
and the subscript referring to subtracting twice on mass shell.

%
%
\subsection{Renormalizing at $q^2=0$}

In this subsection we would like to relate the physical
renormalization discussed above by conventional renormalization at
the unphysical value $q^2=0$ which is convenient for evaluating
integrals.   For the inverse $\phi$ propagator we now expand
$\Sigma(q^2)$ around the point $q^2=0$. That is, we let
\begin{equation}
   \Sigma(p^2) \ \equiv \ \Sigma_{0(0)} \ + \ \Sigma_{1(0)} \ p^2
   \ + \ \Sigma_0^{\mathrm{[sub\, 2]}}(p^2)
   \>,
\end{equation}
where now $\Sigma_{0(0)} = \Sigma(p^2=0)$,  $\Sigma_{1(0)} = \frac
{\mathrm{d}\Sigma}{\mathrm{d}p^2} |_{(p^2=0)}$, and
$\Sigma_0^{\mathrm{[sub\, 2]}}(p^2)$ is again naively finite, but
vanishes quadratically at $p^2=0$. In this case we identify
\begin{equation}
   Z_{2(0)}^{-1} \ \equiv \ - \frac{\mathrm{d}G^{-1}(p^2)}{\mathrm{d}p^2} \Bigr |_{p^2=0}
   \>.
\end{equation}
We can write
\begin{equation}
   G^{-1}(p^2) \ = \ Z_{2(0)}^{-1}  \left[ - p^2 + m_0^2
   \ + \ Z_{2(0)} \ \Sigma_0^{\mathrm{[sub\, 2]}}(p^2)
   \right]
   \>,
\end{equation}
so that
\begin{equation}
   G_R^{-1}(p^2) \ = \
   - p^2 + m_0^2
   \ + \ Z_{2(0)} \ \Sigma_0^{\mathrm{[sub\, 2]}} (p^2)
   \>.
\end{equation}
The same argument as before allows us to symmetrize the dependence
of $\Sigma$ on the vertex function and obtain:
\begin{equation}
   G_R^{-1}(p^2) \ = \ - p^2 + m_0^2 \ + \ \Sigma_{0\, R}^{\mathrm{[sub\, 2]}}(p^2)
   \>.
\end{equation}
The finite mass $m_0^2$ is related to the physical mass which is
the zero of the renormalized Green's function via
\begin{equation}
   0 \ = \ - m^2 + m_0^2 \ + \ \Sigma_{0\, R}^{\mathrm{[sub\, 2]}}(m^2)
   \>.
\end{equation}
This can also be written as
\begin{equation}
   m^2 \ = \ Z_{2(0)}^{-1} \ m_0^2 \ + \ \Sigma(m^2) - \Sigma(0)
   \>.
\end{equation}
For the vertex renormalization we now have
\begin{equation}
   Z_{1(0)}^{-1} \ = \ \Gamma(p,p) |_{p^2=0 }
   \ \equiv \
   1 \ +\ \frac  {\partial \Sigma(p^2)}{\partial \chi} \Bigr |_{p^2=0}
   \>,
\end{equation}
and the Ward-like identity is now $Z_{1(0)} = Z_{2(0)}$.

%
%
\section{Renormalized S-D equations up to order~1/N}
\label{sec:1overN}

Now that we have an all orders renormalization procedure, to
renormalize these equations at order~1/N we realize that $\Delta
\Gamma$ is of order~1/N so it can be ignored in the integral
equations for $\Pi$ and $\Sigma$. Thus, it is consistent with
the~1/N approximation to set $\Gamma_R =1$.  This now gives us the
finite renormalized equations for the inverse Green's functions,
exact up to order~1/N. We can now write the renormalized S-D
equations in the unbroken symmetry case as follows:

For the $\phi$ inverse propagator we have  ($G_{ij} = G \,
\delta_{ij}) $:
\begin{equation}
   G^{-1}_R (q^2) \ = \ m^2 - p^2 \ + \ \Sigma_R^{\mathrm{[sub\, 2]}}(p^2)
   \>.
\end{equation}
with
\begin{equation}
   \Sigma_R (x,y) \ = \ i \, G_R (x,y)  D_R(x,y)
   \>,
\end{equation}
so that in momentum space
\begin{equation}
   \Sigma_R(q^2) \ = \ i \ \int [d^dp] \ G_R(q-p) D_R(p)
   \>.
\end{equation}
We have
\begin{align}
   \Sigma_R^{\mathrm{[sub\, 2]}}(p^2) \ = \ &
   \Sigma_{R}(p^2) \ - \ \Sigma_{R}(p^2=m^2)
   \notag \\ &
   \ - \
   (p^2-m^2) \ \Sigma_{1\, R}(p^2=m^2)
   \>.
   \label{eq:sig2r}
\end{align}
For the $\chi$ inverse propagator we have ($g_R = \lambda_R/N$)
\begin{equation}
   D_R^{-1}(q^2) \ = \ - \ g_R^{-1} \ + \ \Pi_R^{\mathrm{[sub\, 1]}}(q^2)
   \>,
\end{equation}
where
\begin{equation}
   \Pi_R^{\mathrm{[sub\, 1]}}(p^2) \ = \ \Pi_R(p^2) - \Pi_R(0)
   \>,
\end{equation}
and
\begin{equation}
  \Pi_R(x,y) \ = \ \frac{iN}{2} \ G_R(x,y) G_R(y,x)
  \>,
\end{equation}
or in momentum space
\begin{equation}
   \Pi_R(q^2) \ = \ \frac{iN}{2} \ \int [d^dp] \ G_R(q-p) G_R(p)
   \>.
\end{equation}
In order to solve  these equations, one first specifies the value
of the renormalized mass, $m^2$, and the renormalized coupling
constant, $g_R$. Then, both $G_R(p^2)$ and $D_R(p^2)$, and
therefore $\Sigma_R(p^2)$, are just functionals of the finite
quantity $\Sigma_R^{\mathrm{[sub\, 2]}}(p^2)$. Then
Eq.~\eqref{eq:sig2r} is the self-consistent equation we need to
solved for the $\Sigma_R^{\mathrm{[sub\, 2]}}(p^2)$.

One way of solving this equation at large-N is to realize that
$\Sigma_R^{\mathrm{[sub\, 2]}}$ is finite and  of order~1/N so
that one can start with the value of $\Sigma_R^{\mathrm{[sub\,
2]}} $ found in leading order in the large-N approximation and
iterate until convergence is obtained. The strategy for doing this
in the related approximation of the keeping the three-loop sunset
graph in $\Gamma_2$ (which is the first term in a coupling
constant reexpansion of the composite field propagator $D$) is very
clearly discussed in the papers of H. Van Hees and
J.Knoll~\cite{ref:VanHees, ref:sunset} and we will essentially
repeat their strategy here.  Once $\Sigma^{\mathrm{[sub\, 2]}}$ is
obtained, then one can reconstruct $G_R$ and $D_R$, and finally
the subtraction terms $\Sigma_R(m^2)$ and $\Sigma_{1R}(m^2)$
needed for the renormalization of the time dependent evolution
equations, as well as the renormalization constants  $Z_1=Z_2$. If
one instead renormalizes at $q^2= 0$, one first specifies the
parameters $m_0^2$ and $g_R$, and then gets a self-consistent
equation to solve for $\Sigma_{0\, R}^{\mathrm{[sub\, 2]}}(p^2)$.
Again one reconstructs the propagators, the self-energy and vacuum
polarization, and determines the physical mass from the parameter
$m_0^2$ using
\begin{equation}
   m^2 \ = \ m_0^2 \ + \ \Sigma_{0\, R}^{\mathrm{[sub\, 2]}}(p^2=m^2)
   \>.
\end{equation}
The advantage of the mass shell renormalization is that one
definitely has a clear separation between the pole contribution to
the dispersion relation for the Green's function discussed below
and the three-particle cut which starts at~$9m^2$.  This large
mass for the cut then allows a rapidly converging iteration
strategy for $\Sigma$ starting with the leading order in large-N
propagators.

%
%
\subsection{General strategy}

Assuming that indeed the Green's functions $G_R(p^2)$ and
$D_R(p^2)$ vanish at infinity (which is true in 3+1 dimensions, but
will need to be modified in $2+1$ dimensions for $D_R(p)$), then
we can write the spectral representations for the renormalized
Green's functions as
\begin{align}
   G_R(p^2) & =
   \frac{1}{m^2 - p^2 + \Sigma_R^{\mathrm{[sub\, 2]}}(p^2) -i \epsilon}
\label{eq:g_def}
   \\ &
   =    \int_0^\infty \frac{\mathrm{d}(m_1^2)}{\pi}
      \frac{\mathrm{Im} G_R(m_1^2)}{m_1^2 - p^2 - i \epsilon}
\label{eq:g_vac}
   \>,
   \\
   D_R(p^2) & =
   \frac{1}{- g_R^{-1} + \Pi_R^{\mathrm{[sub\, 1]}}(p^2)}
\label{eq:d_def}
   \\ &
   =    \int_0^\infty \frac{\mathrm{d}(m^2)}{\pi}
      \frac{\mathrm{Im} D_R(m_2^2)}{m_2^2 - p^2 - i \epsilon}
\label{eq:d_vac}
   \>,
\end{align}
where
\begin{align}
   \Sigma_R(p^2) \ = \ &
   i \int [ \mathrm{d}^d q ] D_R(p-q) G_R(q)
\label{eq:sig_def}
   \\
   \ = \ &
   \int_0^\infty \frac{\mathrm{d}(m_1^2)}{\pi}    \mathrm{Im} D_R(m_1^2) \label{eq:sig_vac}
   \\ \notag & \times
   \int_0^\infty \frac{\mathrm{d}(m_2^2)}{\pi}    \mathrm{Im} G_R(m_2^2)
      K^{[d]}(p^2;m_1^2,m_2^2)
   \>,
   \\
   \Pi_R(p^2) \ = \ &
   \frac{i}{2}    \int [ \mathrm{d}^d q ] G_R(p-q) G_R(q)
\label{eq:pi_def}
   \\
   \ = \ &
   \frac{1}{2}    \int_0^\infty \frac{\mathrm{d}(m_1^2)}{\pi}    \mathrm{Im} G_R(m_1^2)
\label{eq:pi_vac}
   \\ \notag & \times
   \int_0^\infty \frac{\mathrm{d}(m_2^2)}{\pi}
   \mathrm{Im} G_R(m_2^2)
   K^{[d]}(p^2;m_1^2,m_2^2)
   \>,
\end{align}
with
\begin{align}
&
   K^{[d]}(p^2;m_1^2,m_2^2)
\notag \\
   & =    i    \int \frac{[ \mathrm{d}^d q]}
               {(m_1^2 - (p-q)^2 - i \epsilon)(m_2^2 - q^2 - i \epsilon)}
\label{eq:Kd}
   \\ &
   =    -    \int \ \frac{[ \mathrm{d}^d q_E ]}
               {(m_1^2 + (p_E-q_E)^2 )(m_2^2 + q_E^2 )}
   \>,
\end{align}
with $p^2 \rightarrow - p_E^2$. The last (Euclidean) integral is
found for example in Itzykson and Zuber~\cite{Zuber} in terms of
the Feynman parametrization, as
\begin{align}
K^{[d]}&(p^2;m_1^2,m_2^2) \ = \ - \, \frac{\Gamma(2-d/2)}{(4 \pi)^{d/2}}
\label{eq:Kd_int}
\\ \notag & \times
 \int_0^1 ~d \alpha \left[ \alpha (1-\alpha) p^2 +
\alpha m_1^2+(1-\alpha) m_2^2  \right]^{d/2-2} \>.
\end{align}

The advantage of writing things this way is now clear, as the
kernel can be evaluated in $d$ dimensions exactly. The finite part
of $\Sigma_R$ can be obtained by performing the subtractions
directly on the kernel at the physical mass.  In 2+1 dimensions
$\Sigma_R$ requires one subtraction corresponding to mass
renormalization, and in 3+1 there are two subtractions related to
both mass and wave function renormalization.  $\Pi$ is finite in
2+1 dimensions and goes to zero at large momentum, which will lead
to the need for using a subtracted dispersion relation for $D_R$
in three dimensions.  In 3+1 dimensions $\Pi$ is $\log$ divergent
so that one needs to subtract $\Pi$ once, which corresponds to
coupling constant renormalization.  These subtractions when done
on the kernel $K$ then automatically lead to a finite analytic (in
the cut plane) expression for the subtracted kernel, which is then
used to determine $\Sigma^{\mathrm{[sub\, 2]}}$ and
$\Pi^{\mathrm{[sub\, 1]}}$ via the spectral representation.

The above equations will be solved iteratively: We start with the
free renormalized  $\phi$  Feynman propagator
\begin{align}
   G_0(s=p^2)
   \ = \ &
   \frac{1}{m^2 - s - i \epsilon}
\label{eq:thisg0}
   \\ \notag
   \ = \ &
   - \
   \frac{\mathcal{P}}{s - m^2}
   \ + \
   i \, \pi \ \delta(s-m^2)
   \>.
\end{align}
(For convenience, we shall drop for now the subscript $R$ in
denoting \emph{renormalized} quantities, and we will ignore the
required subtractions for the polarization and self-energy.)
Equation~\eqref{eq:thisg0}, in conjunction with
Eq.~\eqref{eq:pi_def}, gives
\begin{align}
   \Pi_0(s) & \ = \
   \frac{i}{2} \
   \int \ [ \mathrm{d}^d q ] \ G_0[(p-q)^2] \ G_0(q^2)
   \notag \\
   & \ = \
   \frac{1}{2} \
   K^{[d]}(s;m^2,m^2)
\label{eq:pi0}
   \>.
\end{align}
Using Eqs.~\eqref{eq:g_vac} and \eqref{eq:sig_def} we obtain, in
order,
\begin{align}
   D_0(s) & \ = \
   \frac{1}{- \, g_R^{-1} + \Pi_0(s)}
   \>,
\end{align}
\begin{align}
   \Sigma_0(s) & \ = \
   \int_0^\infty \ \frac{\mathrm{d}(m_1^2)}{\pi} \
   \mathrm{Im} \ D_0(m_1^2) \
   K^{[d]}(s;m_1^2,m^2)
   \>.
\end{align}

\begin{figure}[t!]
   \includegraphics[width=3.28in]{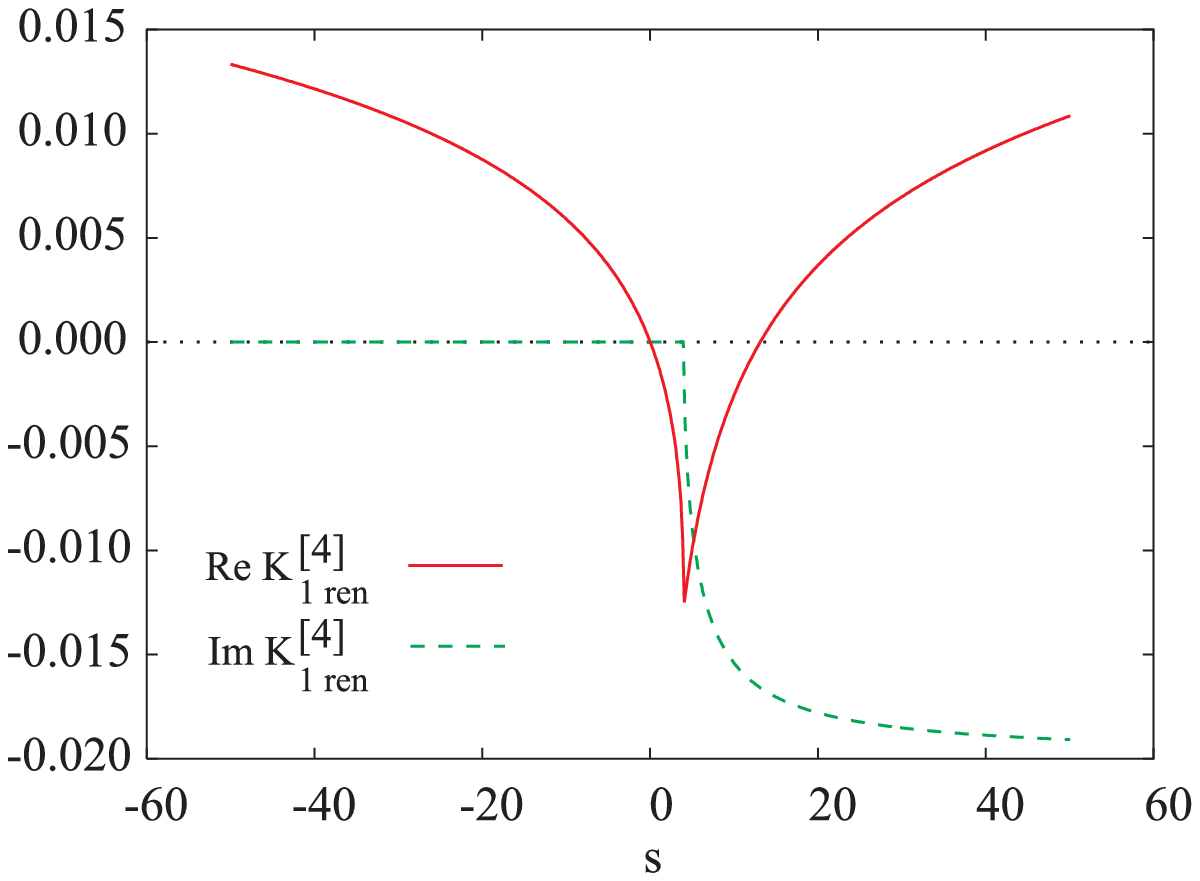}
   \caption{\label{fig:K41}
            (Color online)
            $K^{[4]}_{1\, \mathrm{ren}}(s;m_1^2,m_2^2)$ with $m_1 = m_2 = m = 1$.}
\end{figure}

\begin{figure}[t!]
   \includegraphics[width=3.28in]{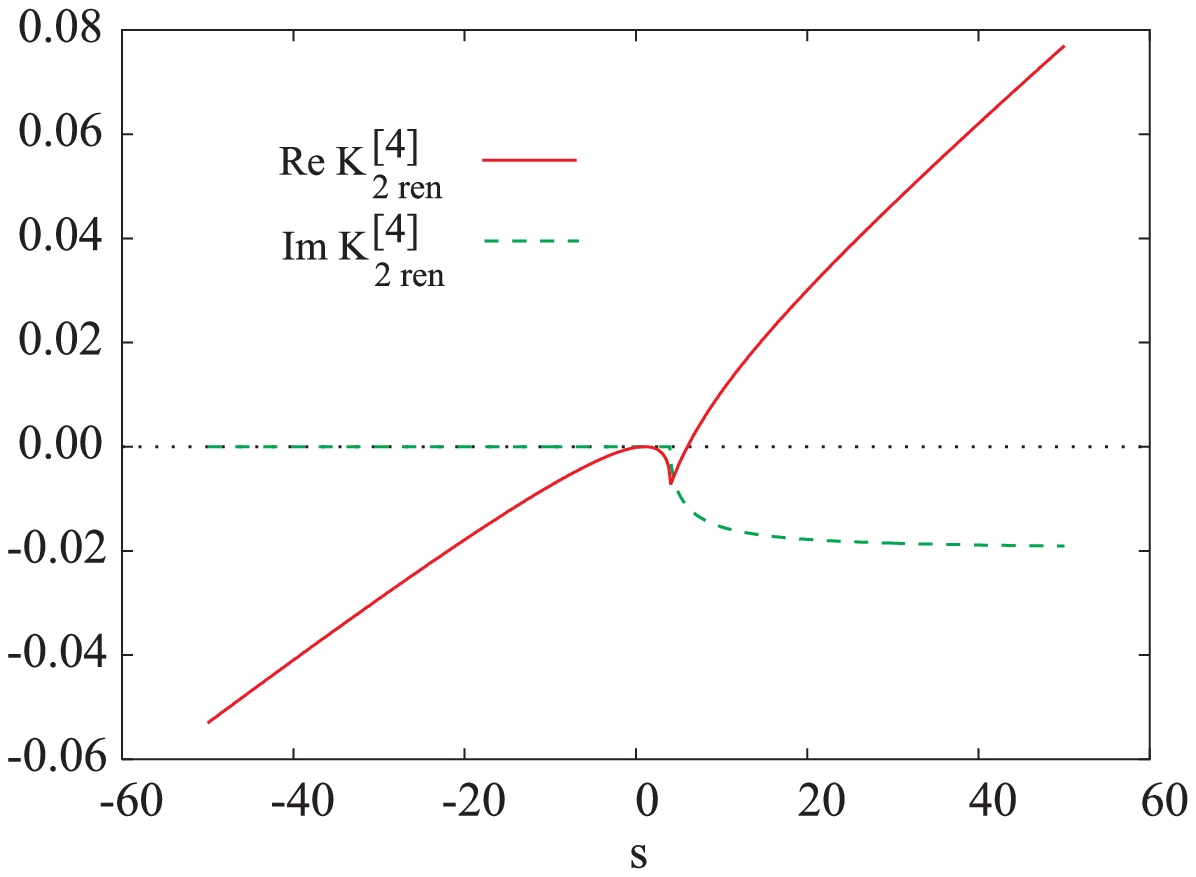}
   \caption{\label{fig:K42}
            (Color online)
            $K^{[4]}_{2\, \mathrm{ren}}(s;m_1^2,m_2^2)$ with $m_1 = m_2 = m = 1$.}
\end{figure}
We can now calculate the first update of the $\phi$ propagator,
$G_1(s)$,
\begin{align}
   G_1(s) & \ = \
   \frac{1}{m^2 - s + \Sigma_0(s)}
   \ = \
   G_0(s)
   \ - \
   \Delta G(s)
\label{eq:G_update}
   \>,
\end{align}
with
\begin{align}
   \Delta G(s)
   \ = \
   \frac{\Sigma_0(s)}{(m^2 - s) \, [m^2  - s + \Sigma_0(s)]}
\label{eq:delG}
   \>.
\end{align}
The updates of the polarization and self-energy are
obtained by combining Eq.~\eqref{eq:G_update} with
Eqs.~\eqref{eq:pi_vac} and \eqref{eq:sig_vac}. We have:
\begin{align}
   \Pi_1(s) \ = \ &   \Pi_0(s)
   -    \int_0^\infty \frac{\mathrm{d}(m_1^2)}{\pi} \,
   \mathrm{Im} \Delta G(m_1^2)    K^{[d]}(s;m_1^2,m^2)
   \notag \\ &
\label{eq:pi1}
   +    \frac{1}{2}    \int_0^\infty \frac{\mathrm{d}(m_1^2)}{\pi} \,
      \mathrm{Im} \Delta G(m_1^2)
   \\ \notag & \qquad \times
         \int_0^\infty \frac{\mathrm{d}(m_2^2)}{\pi} \,
            \mathrm{Im} \Delta G(m_2^2) K^{[d]}(s;m_1^2,m_2^2)
   \>,
\end{align}
\begin{align}
\label{eq:sig1}
   \Sigma_1(s) \ = \ &   \Sigma_0(s)
   -    \int_0^\infty \frac{\mathrm{d}(m_1^2)}{\pi} \, \mathrm{Im} D_1(m_1^2)
      \\ \notag & \qquad \times
   \int_0^\infty \frac{\mathrm{d}(m_2^2)}{\pi} \, \mathrm{Im} \Delta G(m_2^2)
      K^{[d]}(s;m_1^2,m_2^2)
   \>,
\end{align}
where
\begin{align}
   D_1(p) & \ = \
   \frac{1}{- \, g^{-1} + \Pi_1(p)}
   \>.
\end{align}
We can now obtain the ``new'' correction $\Delta G(s)$, by
replacing $\Sigma_0(s)$ with $\Sigma_1(s)$ in Eq.~\eqref{eq:delG}.
Then, we iterate Eqs.~\eqref{eq:delG}, \eqref{eq:pi1}
and~\eqref{eq:sig1}, until we achieve convergence.

\begin{figure*}[t!]
   \includegraphics[width=7.2in]{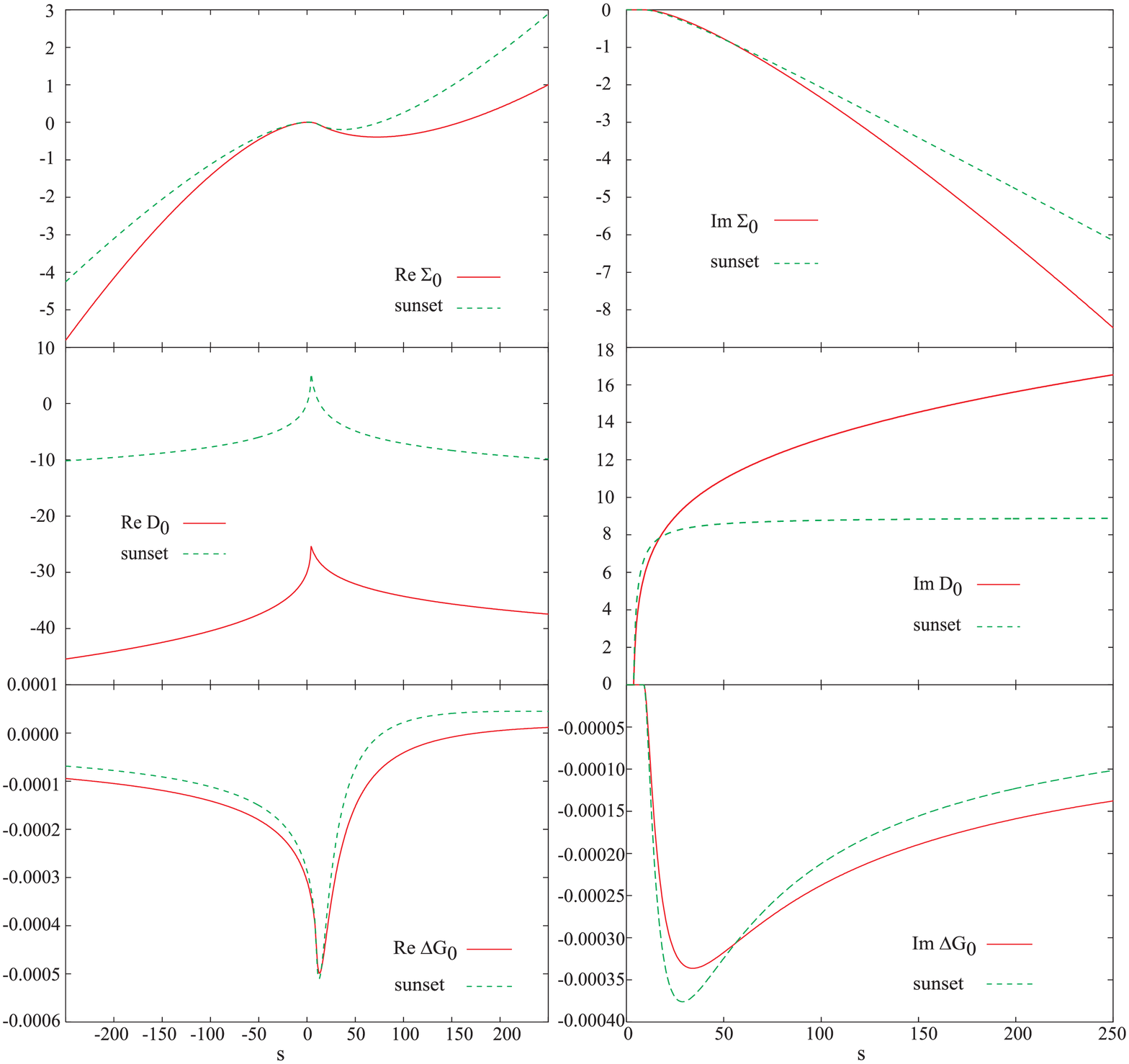}
   \caption{\label{fig:4d}
            (Color online)
            First iteration results in 3+1 dimensions ($g_R$=30).}
\end{figure*}
%
%

\subsection{3+1 dimensions}

The key ingredients for this case have been previously discussed
in van Hees and Knoll~\cite{ref:sunset}. Those authors have used
dimensional regularization arguments in order to regularize the
kernel
\begin{align}
   K^{[4]}&(s=p^2,m_1^2,m_2^2) \ = \
   \\ \notag &
      i    \int [ \mathrm{d}^{4} q ]
           \frac{\mu^{2\epsilon}}
             {[m_1^2-(p-q)^2-i\epsilon](m_2^2-q^2-i\epsilon)}
   \>.
\end{align}
In the range $(m_1-m_2)^2 \le s < (m_1+m_2)^2$, one obtains
\begin{align}
   K^{[4]}& ( s, m_1^2, m_2^2 )
   =
   \frac{1}{16 \pi^2}
   \biggl \{
      -
      \frac{1}{\epsilon}
      +
      \gamma
      -
      2
      \ + \
      \ln
      \Bigl [
         \frac{m_1 m_2}{4 \pi  \mu^2}
      \Bigr ]
   \notag \\ &
      + \
      \frac{ m_1^2 -  m_2^2 }{s} \
      \ln \Bigl [  \frac{m_1}{m_2}  \Bigr ]
      \ + \
      \frac{\bar \lambda(s,m_1^2,m_2^2)}{s}
   \notag \\ & \quad \times
      \biggl [
         \tan^{-1}
         \Bigl [
            \frac{m_{+} m_{-} + s}{\bar \lambda(s,m_1^2,m_2^2)}
         \Bigr ]
         -
         \tan^{-1}
         \Bigl [
            \frac{m_{+} m_{-} - s}{\bar \lambda(s,m_1^2,m_2^2)}
         \Bigr ]
      \biggr ]
   \notag \\ &
      + \
      \mathcal{O}( \epsilon )
      \ + \
      \dotsb
   \biggr \}
\label{e:dr.Kvii}
   \>,
\end{align}
where $\gamma$ is the Euler-Mascheroni constant, and we have
introduced the notation
\begin{align}
   \bar \lambda(s,m_1^2,m_2^2)
   & =
   \sqrt{ - \ \bigl [ s - ( m_1 + m_2 )^2 \bigr ]
          \bigl [ s - ( m_1 - m_2 )^2 \bigr ] }
   \>.
\end{align}

In 3+1 dimensions, both the polarization, $\Pi(s)$, and
self-energy, $\Sigma(s)$, require renormalization. As advertised,
the required prescriptions are
\begin{align}
   \Pi^{\mathrm{[sub\, 1]}}(0) \ = \ & 0 \>,
   \\
   \Sigma^{\mathrm{[sub\, 2]}}(m^2) \ = \ &
   \partial_s \Sigma^{\mathrm{[sub\, 2]}}(s)|_{s=m^2} \ = \ 0 \>,
\end{align}
or
\begin{align}
   \Pi^{\mathrm{[sub\, 1]}}(s) \ \equiv \ &
   \Pi(s) \ - \ \Pi(0)
   \>,
   \\
   \Sigma^{\mathrm{[sub\, 2]}}(s) \ \equiv \ &
   \Sigma(s) \ - \ \Sigma(m^2)
   \ - \ \partial_s \Sigma(s) |_{s = m^2}
   \>.
\end{align}
As such, the renormalized polarization $\Pi^{\mathrm{[sub\,
1]}}(s)$ and self-energy $\Sigma^{\mathrm{[sub\, 2]}}(s)$, will be
calculated in terms of the kernels,
\begin{align}
   K^{[4]}_{1\, \mathrm{ren}}(s;m_1^2,m_2^2)
   & =    K^{[4]}(s;m_1^2,m_2^2)
   -    K^{[4]}(0;m_1^2,m_2^2)
   \>,
\end{align}
and
\begin{widetext}
\begin{align}
   K^{[4]}_{2\, \mathrm{ren}}(s;m_1^2,m_2^2)
   & =    K^{[4]}(s;m_1^2,m_2^2)
   -    K^{[4]}(m^2;m_1^2,m_2^2)
   -    (s - m^2)    \partial_s K^{[4]}(s;m_1^2,m_2^2) \bigr |_{s = m^2}
   \notag \\ &
   =    K^{[4]}_{1\, \mathrm{ren}}(s;m_1^2,m_2^2)
   -    K^{[4]}_{1\, \mathrm{ren}}(m^2;m_1^2,m_2^2)
   -    (s - m^2)    \partial_s K^{[4]}_{1\, \mathrm{ren}}(s;m_1^2,m_2^2) \bigr |_{s = m^2}
   \>.
\end{align}
We obtain
\begin{align}
   K^{[4]}_{1\, \mathrm{ren}}(s;m_1^2,m_2^2)
   \ = \ &
   \frac{1}{16 \pi^{2} s}
   \biggl \{
      - \ s
      \ + \
      \frac{(m_1^2 - m_2^2)^2 \ - \ s \, (m_1^2 + m_2^2)}
           {m_1^2 - m_2^2} \
      \ln \left ( \frac{m_1}{m_2} \right)
\label{K4_1}
   \\ \notag & \qquad
      \ + \
      \bar \lambda(s,m_1,m_2) \
      \biggl [
         \tan^{-1} \frac{s + m_1^2 - m_2^2}
                              {\bar \lambda(s,m_1^2,m_2^2)}
         \ + \
         \tan^{-1} \frac{s - m_1^2 + m_2^2}
                              {\bar \lambda(s,m_1^2,m_2^2)}
      \biggr ]
   \biggr \}
   \>,
\end{align}
and
\begin{align}
   &
   K^{[4]}_{2\, \mathrm{ren}}(s;m_1^2,m_2^2)
   =
   \frac{(m_1^2 - m_2^2) \, (s - m^2)^2}
       {16 \pi^2 M^4 s} \
   \ln \left ( \frac{m_1}{m_2} \right )
\label{K4_2}
   \\ \notag &
   +
   \frac{1}{16 \pi^2 m^2 s}
   \bigg \{
      m^2 \, \bar \lambda(s,m_1,m_2)
      \bigg [
         \tan^{-1} \frac{s + m_1^2 - m_2^2}
                              {\bar \lambda(s,m_1,m_2)}
         +
         \tan^{-1} \frac{s - m_1^2 + m_2^2}
                              {\bar \lambda(s,m_1,m_2)}
      \bigg ]
   \\ \notag & \qquad \qquad \qquad \quad
      -
      s \, \bar \lambda(m^2,m_1,m_2)
      \bigg [
         \tan^{-1} \frac{m^2 + m_1^2 - m_2^2}
                         {\bar \lambda(m^2,m_1,m_2)}
         +
         \tan^{-1} \frac{m^2 - m_1^2 + m_2^2}
                         {\bar \lambda(m^2,m_1,m_2)}
     \bigg ]
  \bigg \}
  \\ \notag &
  -
  \frac{s - m^2}{16 \pi^2 m^2}
  \bigg \{
     1
     +
     \frac{(m_1^2 - m_2^2)^2 - m^2 \, (m_1^2 + m_2^2)}
          { \bar \lambda(m^2,m_1,m_2) m^2 }
     \bigg [
        \tan^{-1} \frac{m^2 + m_1^2 - m_2^2}{\bar \lambda(m^2,m_1,m_2)}
        +
        \tan^{-1} \frac{m^2 - m_1^2 + m_2^2}{\bar \lambda(m^2,m_1,m_2)}
     \bigg ]
  \bigg \}
  \>.
\end{align}
Equations \eqref{K4_1} and \eqref{K4_2} are analytically continued
outside the range $(m_1-m_2)^2 \le s < (m_1+m_2)^2$, using
\begin{align}
   \tan^{-1} \frac{s + m_1^2 - m_2^2}{\bar \lambda(s,m_1,m_2)}
   &
   +
   \tan^{-1} \frac{s - m_1^2 + m_2^2}{\bar \lambda(s,m_1,m_2)}
   \\ \notag &
   =
\left \{
\begin{array}{ll}
   \frac{1}{i}
   \ln \frac{ \sqrt{(m_1+m_2)^2 - s} - \sqrt{(m_1-m_2)^2 - s} }
            { \sqrt{(m_1+m_2)^2 - s} + \sqrt{(m_1-m_2)^2 - s} }
   \>,
   &
   \mathrm{if} \
   s \leq (m_1 - m_2)^2
   \>,
   \\
   - \
   \frac{1}{i}
   \Bigl [
   \ln \frac{ \sqrt{s - (m_1-m_2)^2} - \sqrt{s - (m_1+m_2)^2} }
            { \sqrt{s - (m_1-m_2)^2} + \sqrt{s - (m_1+m_2)^2} }
   \ + \
   i \pi
   \Bigr ]
   \>,
   &
   \mathrm{if} \
   s > (m_1 + m_2)^2
   \>.
\end{array}
\right .
\end{align}
\end{widetext}
We note that the final renormalized expression for
$K^{[4]}_{1ren}$ could just as well have been obtained from
covariant or non-covariant cutoff methods of regularization.

For illustrative purposes, in Figs.~\ref{fig:K41} and
\ref{fig:K42}, we depict the $s$-dependence of $K^{[4]}_{1\,
\mathrm{ren}}(s;m^2,m^2)$ and $K^{[4]}_{1\,
\mathrm{ren}}(s;m^2,m^2)$. (We recall that $K^{[4]}_{1\,
\mathrm{ren}}(s;m^2,m^2)$ is related to the starting value of the
polarization, $\Pi_{0}(s)$, see Eq.~\eqref{eq:pi0}.)

\begin{figure}[b!]
   \includegraphics[width=3.3in]{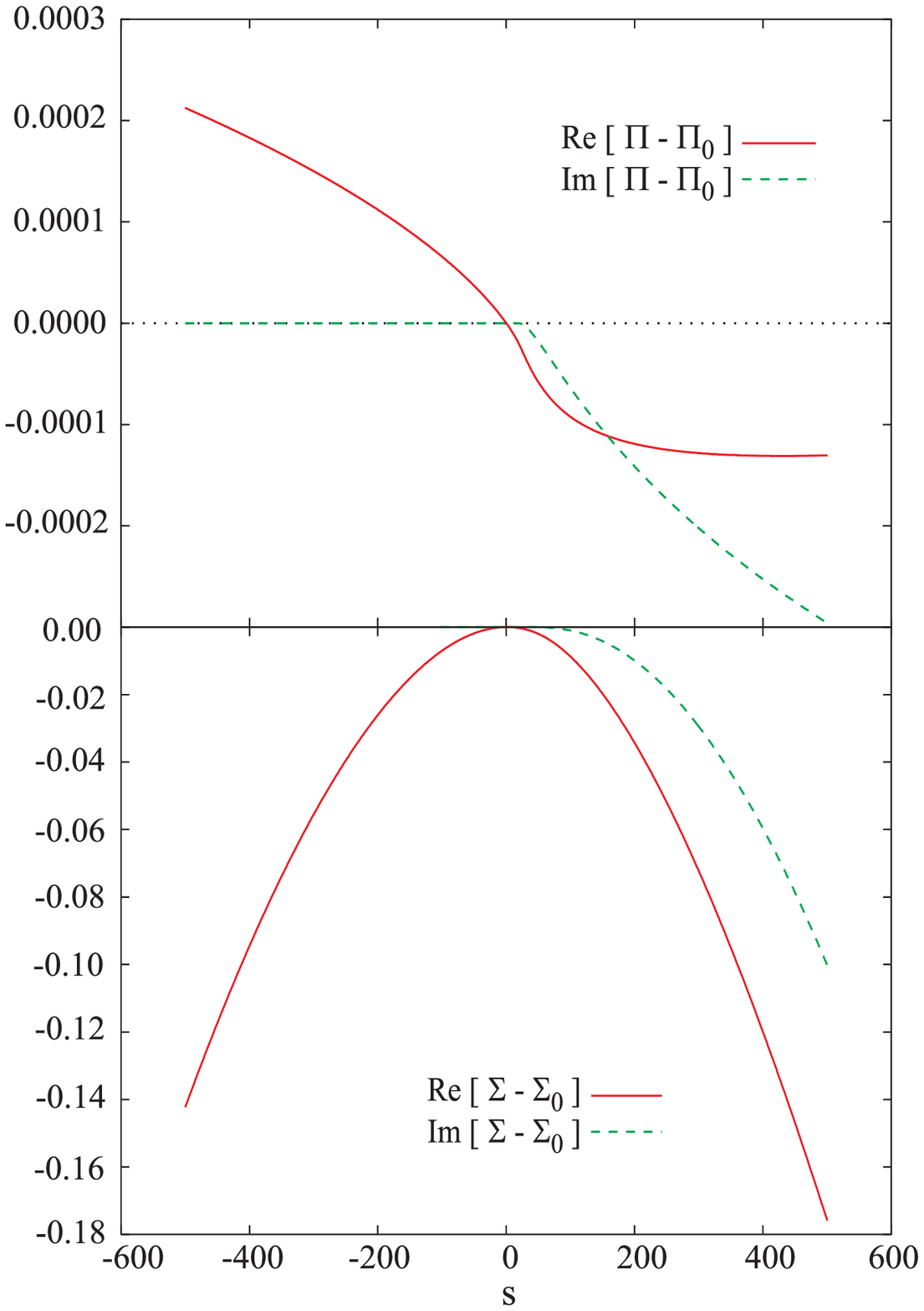}
   \caption{\label{fig:4d_diff}
            (Color online)
            Self-consistent values of the polarization and self-energy in 3+1 dimensions.}
\end{figure}

\begin{figure}[t!]
   \includegraphics[width=3.47in]{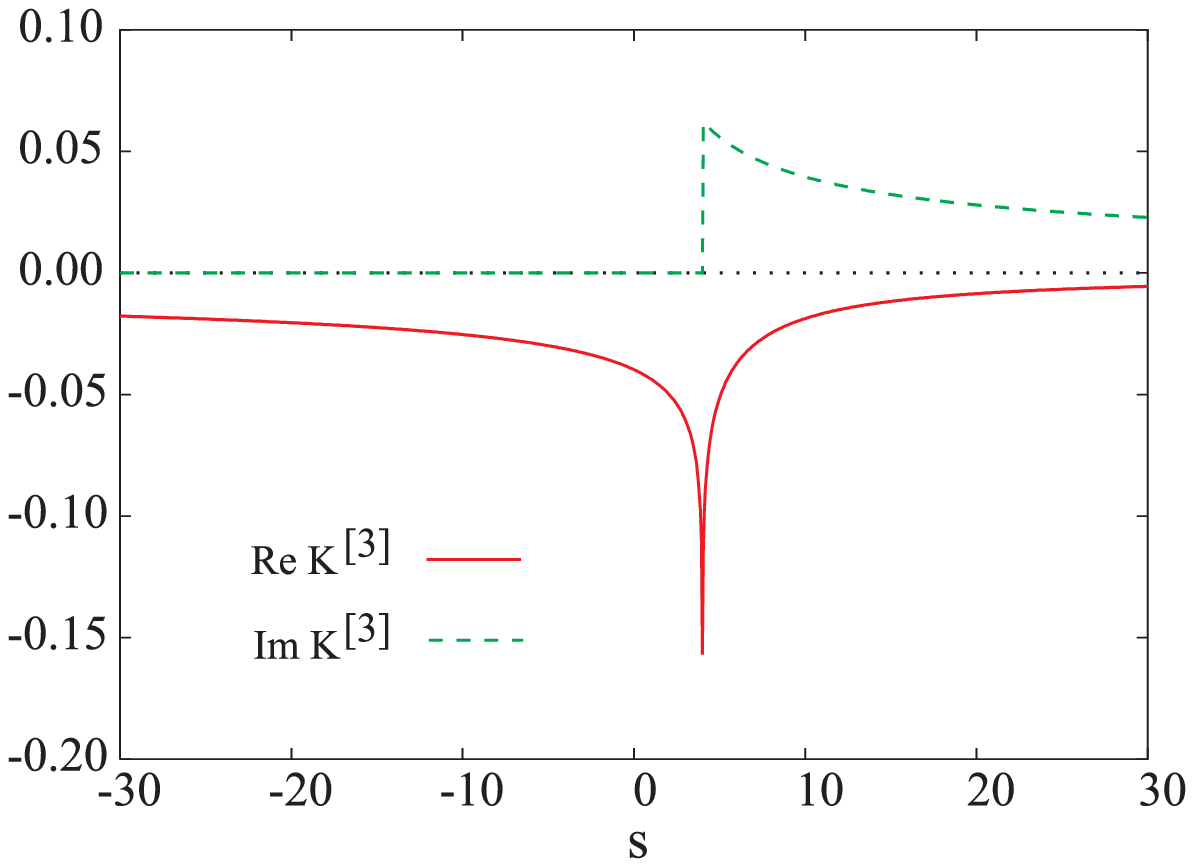}
   \caption{\label{fig:K31}
            (Color online)
            $K^{[3]}(s;m_1^2,m_2^2)$ with $m_1 \!=\! m_2 \!=\! m \!=\! 1$.}
\end{figure}

\begin{figure}[t!]
   \includegraphics[width=3.47in]{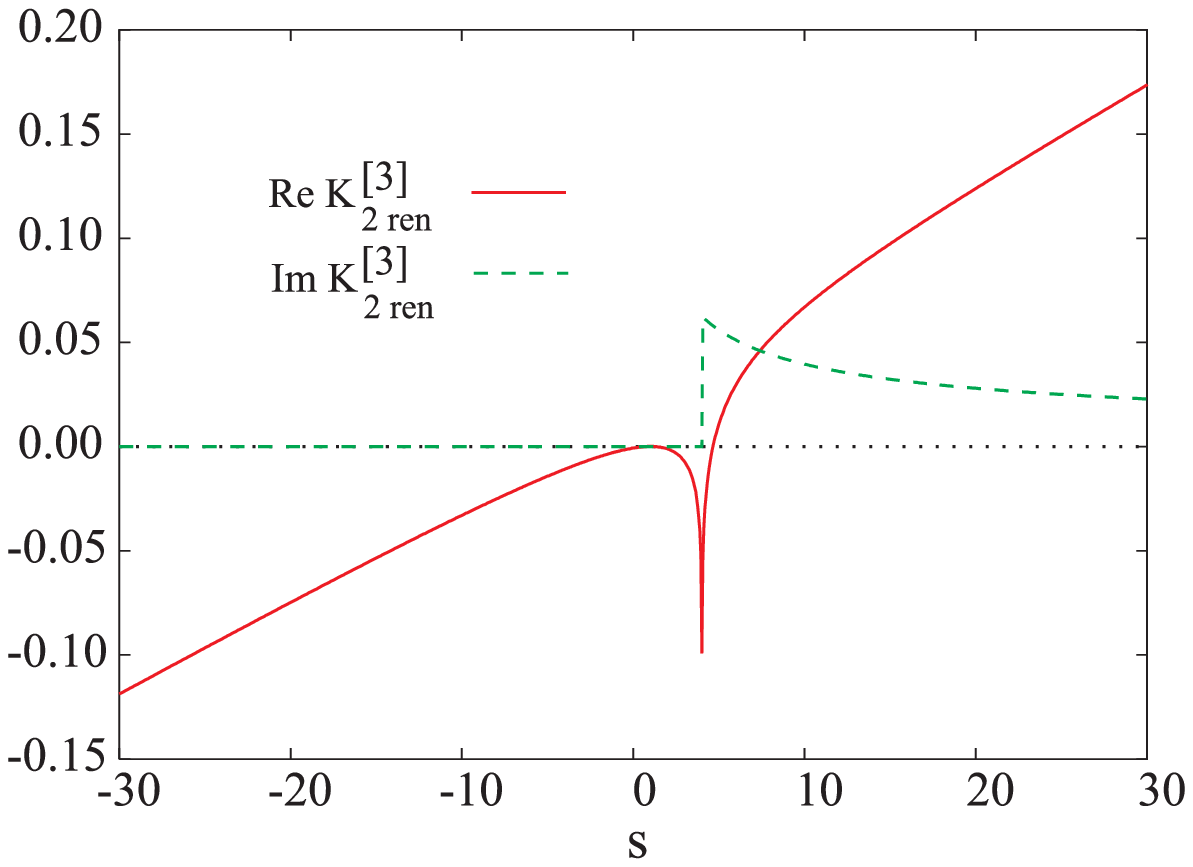}
   \caption{\label{fig:K32}
            (Color online)
            $K^{[3]}_{2\, \mathrm{ren}}(s,m;m_1^2,m_2^2)$ with $m_1 = m_2 \!=\! m \!=\! 1$.}
\end{figure}

We begin our iterations by calculating
\begin{align}
   G_0(s) & \ = \
   \frac{1}{m^2 - s - i \epsilon}
   \>,
\end{align}
\begin{align}
   \Pi^{\mathrm{[sub\, 1]}}_{0}(s) & \ = \
   \frac{1}{2} \
   K^{[4]}_{1\, \mathrm{ren}}(s;m^2,m^2)
   \>,
\end{align}
\begin{align}
   D_0(s) & \ = \
   \frac{1}{- \, g_R^{-1} \ + \ \Pi^{\mathrm{[sub\, 1]}}_{0}(s)}
   \>,
\end{align}
\begin{align}
   \Sigma^{\mathrm{[sub\, 2]}}_{0}(s) & \ = \
   \int_{4m^2}^\infty \ \frac{\mathrm{d}(m_1^2)}{\pi} \,
   \mathrm{Im} \, D_0(m_1^2) \,
   K^{[4]}_{2\, \mathrm{ren}}(s;m_1^2,m^2)
   \>.
\end{align}
Subsequently we iterate
\begin{align}
   G_1(s) = &
   \frac{1}{m^2 - s + \Sigma^{\mathrm{[sub\, 2]}}_{0}(s)}
   =    G_0(s)
   -    \Delta G(s)
   \>,
\end{align}
\begin{align}
   \Pi^{\mathrm{[sub\, 1]}}_1&(s) =
   \Pi^{\mathrm{[sub\, 1]}}_0(s)
   \\ \notag &
   -    \int_{9m^2}^\infty \ \frac{\mathrm{d}(m_1^2)}{\pi} \
   \mathrm{Im} \Delta G(m_1^2) \ K^{[4]}_{1\, \mathrm{ren}}(s;m_1^2,m^2)
   \\ \notag &
   +    \frac{1}{2}    \int_{9m^2}^\infty \ \frac{\mathrm{d}(m_1^2)}{\pi} \
   \mathrm{Im} \Delta G(m_1^2)
   \\ \notag & \qquad \times
   \int_{9m^2}^\infty \ \frac{\mathrm{d}(m_2^2)}{\pi} \,
   \mathrm{Im} \Delta G(m_2^2)    K^{[4]}_{1\, \mathrm{ren}}(s;m_1^2,m_2^2)
   \>,
\end{align}
\begin{align}
   D_1(s) \ = \ &
   \frac{1}{- \, g_R^{-1} \ + \ \Pi^{\mathrm{[sub\, 1]}}_1(s)}
   \>,
\end{align}
\begin{align}
   \Sigma^{\mathrm{[sub\, 2]}}_{1}&(s) =
   \Sigma^{\mathrm{[sub\, 2]}}_0(s)
   -    \int_{4m^2}^\infty \frac{\mathrm{d}(m_1^2)}{\pi} \,
      \mathrm{Im} D_1(m_1^2)
         \\ \notag & \times
   \int_{9m^2}^\infty \ \frac{\mathrm{d}(m_2^2)}{\pi} \
      \mathrm{Im} \Delta G(m_2^2) \
         K^{[4]}_{2\, \mathrm{ren}}(s;m_1^2,m_2^2)
   \>.
\end{align}

The first iteration results are depicted in Fig.~\ref{fig:4d}.
Here, we compare our results with the leading order contribution
in the perturbative reexpansion of $D(s)$ in powers of $\Pi(s)$,
i.e.
\begin{align}
   D(s) & \ = \
   \frac{1}{- \, g_R^{-1} + \Pi(s)}
   \\ \notag &
   \ = \
   - \ g_R \ - \ g_R^2 \ [\Pi^{\mathrm{[sub\, 1]}}(s)]^2 \ - \ \cdots
   \>,
\end{align}
which is similar to the \emph{sunset} approximation of Van Hees
and Knoll.

The effect of the self-consistent calculation is depicted in
Fig.~\ref{fig:4d_diff}. As stated by van Hees and
Knoll~\cite{ref:sunset}, the first iteration results are
negligibly modified by the subsequent iterations, since the main
contributions come from the pole term of $G_0$, and the continuous
corrections start at a threshold of $s=9m^2$. Henceforth, only the
high momentum tail of the propagators is affected, and convergence
is rapidly achieved.

%
%

\subsection{2+1 dimensions}

The 2+1 dimensions case is different from the 3+1 dimensions case
for two reasons. The first is that the vacuum polarization goes to
zero at large momentum so that $D$ goes to a constant and obeys a
once-subtracted dispersion relation.  Secondly, the self-energy
requires only one subtraction to make it finite. However, we will
find it more convenient for our iterative scheme to use two
subtractions so that the renormalized Green's function has the
{\it same} strength at the pole as the free one.

\begin{figure*}[t!]
   \includegraphics[width=6.9in]{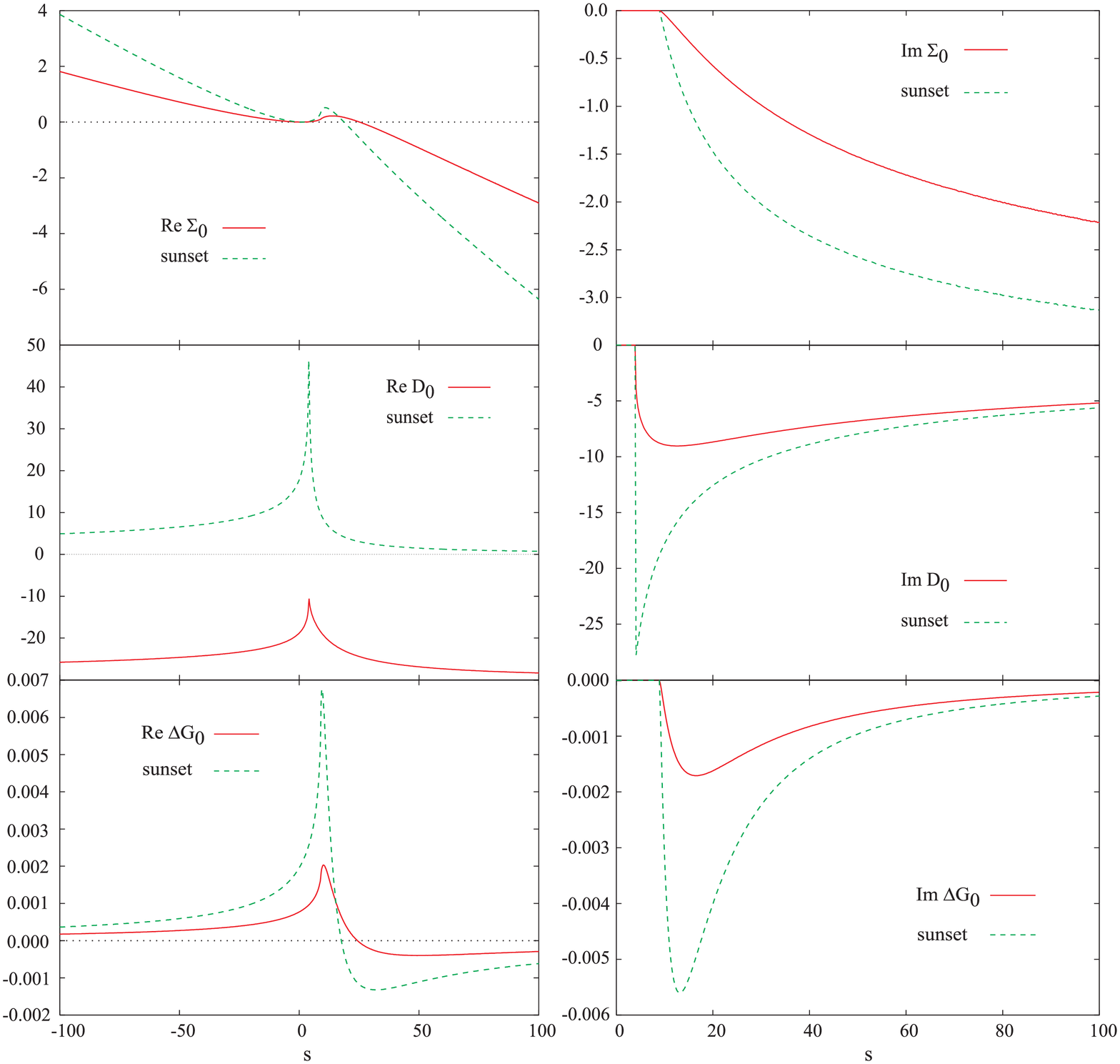}
   \caption{\label{fig:3d}
            (Color online)
            First iteration results in 2+1 dimensions ($g$=30).}
\end{figure*}

Using Eq.~\eqref{eq:Kd_int}, the kernel $K^{[3]}(s;m_1^2,m_2^2)$
is evaluated for the domain $0 < s \le (m_1 + m_2)^2$, and the
result is analytically continued outside this range. We obtain
\begin{widetext}
\begin{align}
   K^{[3]}(s;m_1^2,m_2^2) & \ = \
\left \{
\begin{array}{ll}
   - \
   \frac{1}{4\pi \, \sqrt{|s|}} \
   \tan^{-1} \frac{\sqrt{|s|}}{m_1 + m_2}
   \>,
   &
   \mathrm{if} \
   s \leq 0
   \>,
   \\
   - \
   \frac{1}{8\pi \, \sqrt{s}} \
   \ln \frac{m_1 + m_2 + \sqrt{s}}{m_1 + m_2 - \sqrt{s}}
   \>,
   &
   \mathrm{if} \
   0 \leq s < (m_1 + m_2)^2
   \>,
   \\
   - \
   \frac{1}{8\pi \, \sqrt{s}} \
   \left [
   \ln \frac{\sqrt{s} + (m_1 + m_2)}{\sqrt{s} - (m_1 + m_2)}
   -
   i \, \pi
   \right ]
   \>,
   &
   \mathrm{if} \
   s \ge (m_1 + m_2)^2
   \>.
\end{array}
\right . \label{eq:k3}
\end{align}
\end{widetext}
In Fig.~\ref{fig:K31} we depict the $s$-dependence of
$K^{[3]}(s;m^2,m^2)$, which is related to the starting value of
the polarization, $\Pi_0(s)$. We notice immediately that for large
momentum, we have $K^{[3]}(s;m_1^2,m_2^2) \rightarrow 0$, like
$1/{\sqrt s}$. In order to satisfy the boundary conditions
required by the spectral representation, we use the simple
manipulation
\begin{align}
   D(s) & \ = \
   - \
   g \ + \
   [ D(s) + g ]
   \\ &
   \ = \
   - \
   g \ + \
   \int_{4m^2}^\infty \ \frac{\mathrm{d}(m^2)}{\pi} \
   \frac{\mathrm{Im} \, D(m^2)}{m^2 - s - i \epsilon}
\label{eq:d_vac_3d}
   \>,
\end{align}
which gives
\begin{align}
&
   \Sigma(s)
   =
   -    i g
   \int [ \mathrm{d}^3 q ] G(q^2)
   +    \int_{4m^2}^\infty \frac{\mathrm{d}(m_1^2)}{\pi}    \mathrm{Im} \, D(m_1^2)
      \notag \\ & \qquad \times
   \int_0^\infty \frac{\mathrm{d}(m_2^2)}{\pi}    \mathrm{Im} G(m_2^2)    K^{[3]}(s;m_1^2,m_2^2)
\label{eq:3d_sig}
   \>.
\end{align}
Thus we see that one subtraction  (mass renormalization) is
sufficient to render the theory finite.
However making only one subtraction, i.e.
\begin{align}
   \Sigma^{\mathrm{[sub\, 1]}}(m^2) & \ = \ 0 \>,
\end{align}
or
\begin{align}
   \Sigma^{\mathrm{[sub\, 1]}}(s) & \ \equiv \
   \Sigma(s) \ - \
   \Sigma(m^2)
   \>,
\end{align}
one induces a finite wave function renormalization making the bare
and renormalized Green's functions having different strengths at
the pole. To avoid this, it is convenient to do a complete
physical renormalization even in 2+1 dimensions and instead
consider:
\begin{align}
   \Sigma^{\mathrm{[sub\, 2]}} (s) = &
      \int_{4m^2}^\infty \frac{\mathrm{d}(m_1^2)}{\pi}    \mathrm{Im} \, D(m_1^2)
   \\ \notag & \quad \times
         \int_0^\infty \frac{\mathrm{d}(m_2^2)}{\pi}    \mathrm{Im} G(m_2^2)
            K^{[3]}_{2\, \mathrm{ren}}(s;m_1^2,m_2^2)
   \>,
\end{align}
where
\begin{align}
   & K^{[3]}_{2\, \mathrm{ren}}(s;m_1^2,m_2^2)
   \, = \,
        K^{[3]}(s;m_1^2,m_2^2)
   -    K^{[3]}(m^2;m_1^2,m_2^2)
   \notag \\ & \qquad
   - (s-m^2) \ \partial_s  K^{[3]}(s;m_1^2,m_2^2) |_{s=m^2}
   \>.
\end{align}
For the range $0 \leq s < (m_1+m_2)^2$, we obtain
\begin{align}
   K^{[3]}_{2\, \mathrm{ren}}&(s;m_1^2,m_2^2)
   \ = \
   - \ \frac{1}{8\pi^2}
   \biggl [
   \frac{1}{\sqrt{s}} \
   \ln \frac{m_1 + m_2 + \sqrt{s}}{m_1 + m_2 - \sqrt{s}}
   \notag \\ &
   \ - \
   \Bigl ( 1 - \frac{s - m^2}{2 m^2} \Bigr ) \
   \ln \frac{m_1 + m_2 + m}{m_1 + m_2 - m}
   \notag \\ &
   \ - \
   \frac{(s-m^2)(m_1+m_2)}{m^2 \bigl [ (m_1+m_2)^2 - m^2 \bigr ]}
   \biggr ]
   \>.
\end{align}
The analytical continuation of the above result is done according
to Eq.~\eqref{eq:k3}, and we illustrate in Fig.~\ref{fig:K32} the
$s$-dependence of $K^{[3]}_{2\, \mathrm{ren}}(s;m^2,m^2)$.

For concreteness, we list the explicit equations we need the
solve. For the first iteration, we have
\begin{align}
   G_0(s) & =    \frac{1}{m^2 - s - i \epsilon}
   \>,
\end{align}
\begin{align}
   \Pi_0(s) & =    \frac{1}{2} \ K^{[3]}(s;m^2,m^2)
   \>,
\end{align}
\begin{align}
   D_0(s) & =    \frac{1}{- \, g^{-1} + \Pi_0(s)}
   \>,
\end{align}
\begin{align}
   \Sigma^{\mathrm{[sub\, 2]}}_{0}(s) & =
      \int_{4m^2}^\infty \frac{\mathrm{d}(m_1^2)}{\pi} \
         \mathrm{Im} D_0(m_1^2) \, K^{[3]}_{2\, \mathrm{ren}}(s;m_1^2,m^2)
   \>,
\end{align}
while the subsequent iterations provide the solution of the system
of equations
\begin{align}
   G_1(s) = &
   \frac{1}{m^2 - s + \Sigma^{\mathrm{[sub\, 2]}}_{0}(s)}
   =    G_0(s)
   -    \Delta G(s)
   \>,
\end{align}
\begin{align}
   \Pi_1&(s) =
   \Pi_0(s)
   \\ \notag &
   -    \int_{9m^2}^\infty \ \frac{\mathrm{d}(m_1^2)}{\pi} \
   \mathrm{Im} \Delta G(m_1^2) \ K^{[3]}(s;m_1^2,m^2)
   \\ \notag &
   +    \frac{1}{2} \ \int_{9m^2}^\infty \
   \frac{\mathrm{d}(m_1^2)}{\pi} \
   \mathrm{Im} \Delta G(m_1^2)
   \\ \notag & \qquad \times
   \int_{9m^2}^\infty \ \frac{\mathrm{d}(m_2^2)}{\pi} \
   \mathrm{Im} \Delta G(m_2^2) \ K^{[3]}(s;m_1^2,m_2^2)
   \>,
\end{align}
\begin{align}
   D_1(s) \ = \ &
   \frac{1}{- \, g^{-1} \ + \ \Pi_1(s)}
   \>,
\end{align}
\begin{align}
   \Sigma^{\mathrm{[sub\, 2]}}_{1}&(s) =
   \Sigma^{\mathrm{[sub\, 2]}}_0(s)
   -    \int_{4m^2}^\infty \frac{\mathrm{d}(m_1^2)}{\pi} \
      \mathrm{Im} D_1(m_1^2)
         \\ \notag & \times
   \int_{9m^2}^\infty \frac{\mathrm{d}(m_2^2)}{\pi} \,
      \mathrm{Im} \Delta G(m_2^2)
         K^{[3]}_{2\, \mathrm{ren}}(s;m_1^2,m_2^2)
   \>.
\end{align}

Similarly to the 3+1 dimensions case, we plot the results after
the first iteration (see Fig.~\ref{fig:3d}), and compare with the
\emph{sunset}-like approximation of van Hees and Knoll. Once
again, we the corrections beyond the first iteration result are
suppressed due to $(m_1+m_2)^2$ threshold in the emergence of the
kernels' imaginary parts, and self-consistent result virtually
lies on top of the first iteration result.

%
%
\begin{figure}[b!]
   \includegraphics[width=3.3in]{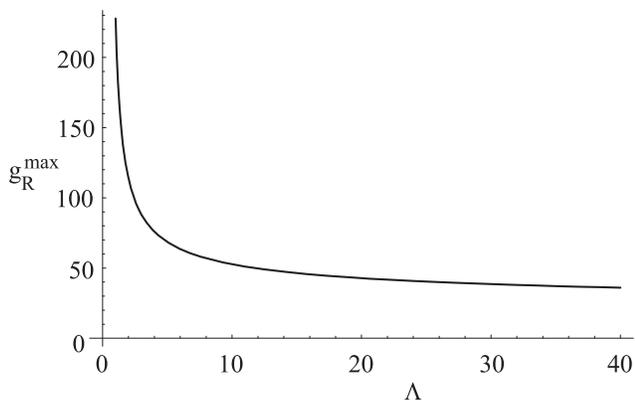}
   \caption{\label{fig:cutoff}
            The cutoff, $\Lambda$, dependence of the maximum value
            of the renormalized coupling constant, $g_R^{\mathrm{max}}$.}
\end{figure}

\section{Effect of Landau Pole}
\label{sec:landau}

What we have done earlier was a slight cheat for 3+1 dimensions in
that $\lambda \phi^4$ field theory is only an {\it effective}
field theory in 3+1 dimensions, having nontrivial scattering only
when defined on the lattice (or with a momentum cutoff), and the
lattice spacing {\it not} taken to zero~\cite{phi4}.   The range
of validity of the effective theory is determined by the position
of the Landau Pole.  The bare coupling constant $ \lambda $  must
be positive for the lattice field theory to be defined. Using
Eq.~\eqref{eq:gtogr}, we obtain the relationship
\begin{equation}
   g_R \ = \
   \frac{g}{1 \ - \ g \, \Pi(0)}\>,
\end{equation}
or
\begin{equation}
   g \ = \
   \frac{g_R}{1 \ + \ g_R \, \Pi(0)}\>,
\end{equation}
If we evaluate $\Pi(0)$ in leading order, integrating over $p_0$
and using a cutoff $\Lambda$ in  $| {\vec p}|$ we have
\begin{equation}
   \Pi(0) \ = \
   - \ \frac {1}{16 \pi^2}
   \int_0^{\Lambda} \frac{p^2 \, \mathrm{d}p}{(p^2+m^2)^{\frac{3}{2}}}
   \>.
\end{equation}
The asymptotic behavior of the above expression at large cutoff,
$\Lambda$, is
\begin{equation}
   \Pi(0) \ = \
   - \frac {1}{16 \pi^2} \
   \ln{\frac{2 \Lambda}{m}}
   \>.
\end{equation}
In the cutoff theory one has that $g_R$ is a monotonically
increasing function of the bare coupling constant $g$, and has a
maximum value defined by
\begin{equation}
   g_R^{\mathrm{max}}
   \ = \
   - \ \frac{1}{\Pi(0)}
   \ = \ 16 \pi^2 \ \Bigl [
\ln \frac{2 \Lambda}{m} \Bigr ]^{-1} \>.
\end{equation}
This behavior is shown in Fig.~\ref{fig:cutoff}.

\begin{figure}[t!]
   \includegraphics[width=3.4in]{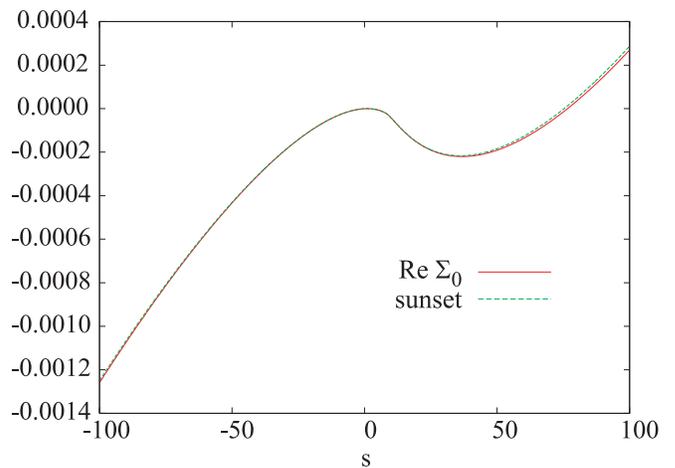}
   \caption{\label{fig:re_sig_g1}
            (Color online)
            Real part of the self-energy $\Sigma_0(s)$, in 3+1
            dimensions, $g_R=1$.}
\end{figure}

\begin{figure}[t!]
   \includegraphics[width=3.4in]{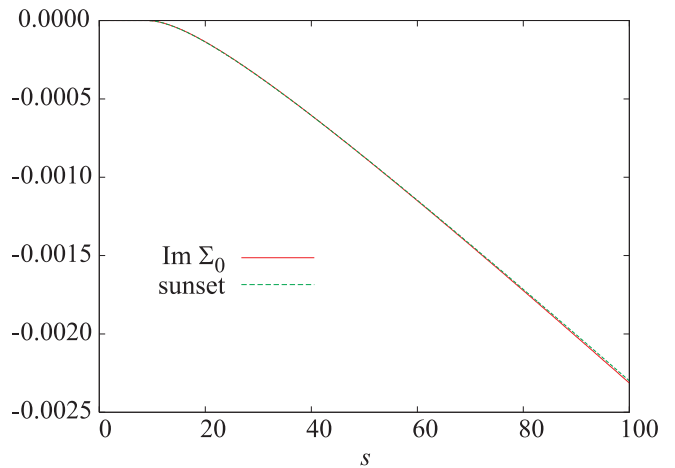}
   \caption{\label{fig:im_sig_g1}
            (Color online)
            Imaginary part of the self-energy $\Sigma_0(s)$, in 3+1
            dimensions, $g_R=1$.}
\end{figure}

In order to capture all the physics of our approximation, we would
like the cutoff, $\Lambda$, to be much larger than the $9m^2$
threshold, that is so important in getting the correct physics.
Thus, any $\Lambda$ greater than say $30m^2$ will be sufficiently
large. From Fig.~\ref{fig:cutoff}, we conclude that as long as
$g_R \sim 1$, there is a wide range of momenta for which the
effective theory is valid, and we can expect there exists a regime
of cutoffs (less than the maximum momentum) for which the theory
becomes cutoff independent. This behavior was shown to be correct
in our mean field simulations of disoriented chiral
condensates~\cite{dcc}. In this regime, the continuum results we
used here should offer a good approximation to the actual cutoff
integrals required for consistency in order that $\phi^4$ field
theory be a good effective field theory in the energy regime lower
than the Landau cutoff for that coupling constant. To avoid Landau
pole issues, using $g_R \leq 1$ is much more realistic than the
value ($g_R$=30) chosen by van Hees and Knoll.  In 3+1 dimensions
at $g_R=1$, the resummed 1/N approximation and the sunset
approximation are indeed very close. These results are illustrated
in Figs.~\ref{fig:re_sig_g1} and~\ref{fig:im_sig_g1}.

%
%
\section{Conclusions}

In this paper we have discussed the renormalization of the S-D
equations of the auxiliary field formulation of $\phi^4$ field theory
and then specialized to the self-consistent approximation to the
coupled Green's function equations obtained by ignoring vertex
corrections or equivalently expanded the 2-PI generating functional in
loops and keeping the two-loop contribution to $\Gamma_2$.  We then
obtained vacuum solutions for the self energy and vacuum polarization
contribution to the inverse Green's functions in the bare vertex
approximation.  We compared our results to the related sunset graph
approximation of Van Hees and Knoll and discovered that at strong
coupling there were significant differences in these two
approximations.  These differences become insignificant when $g_R$ is
of order 1. In discussing the renormalization we confined ourselves to
the case when there is no symmetry breakdown.  When there is symmetry
breakdown, one can use a ``mass independent'' renormalization scheme,
such as that proposed by Kugo~\cite{ref:Kugo}, which requires only
renormalizing the effective action in the symmetric theory, which is
discussed here, to also renormalize the broken symmetry case.
Otherwise, one can study the more complicated set of broken symmetry
Dyson equations and renormalize those as discussed in
Ref.~\cite{Haymaker1}.  The reason we chose physical on mass shell
renormalization in this paper was that it allowed a clean iterative
method for solving the self consistent S-D equations for the vacuum
sector.  For non-equilibrium uses of the effective action in the CTP
formalism one does not need to rely on self consistent solutions,
since the time update equations will find these solutions dynamically.
In the latter case it might be more efficient to determine the
renormalization counter terms using the mass independent strategy of
Kugo.

An important question we have not addressed here is the fulfilment of
Ward identities.  In obtaining the renormalized S-D equations, we
relied on the Ward identity $Z_1=Z_2$ to obtain our final result for
the exact renormalized Green functions.  After doing that we noticed
it was consistent up to order $1/N$ to set the ``renormalized'' vertex
function to 1.  In leading order (and beyond) in traditional large-N
expansions of the auxiliary field generating functional, one
explicitly can verify (as well as prove order by order) that
$Z_1=Z_2$, and also the Ward identities related to the O(N) symmetry,
are satisfied order by order.  However, the resummation inherent in
the self consistently determined 2-PI effective action can violate the
exact Ward identities since one now has a result exact to a particular
order in $1/N$, but different quantities (for example $Z_1$ and $Z_2$)
may differ in the coefficients of the $1/N^2$ and higher terms when
the relevant Green functions are reexpanded in a series in $1/N$.
Therefore, although we have obtained finite renormalized S-D
equations, which are exact to order $1/N$ one cannot rule out order
$(1/N^2)$ violations of Ward identities, Goldstone theorems etc.  To
obtain approximations in the 2-PI formalism which are consistent with
Ward identities and gauge invariance (for gauge theories) is still an
unsolved problem and the subject of lots of recent
attention~\cite{ref:VanHees66,ward}.  This subject is beyond the scope
of this paper, which is mainly concerned with obtaining finite
renormalized S-D equations for the truncated hierarchy of Green
functions.  In a future publication, we will use the results presented
here to study the thermalization of renormalized $\phi^4$ field theory
in 2+1 and 3+1 dimensions for both $N=1$ and the $N=4$ O(N) model and
study the grid sizes needed for the renormalized theory to be
independent of grid size.

%
%
\begin{acknowledgments}

We would like to thank the Santa Fe Institute for its hospitality
during the completion of this work. We would also like to thank
J.~Berges for useful comments.

\end{acknowledgments}

\vfill

%
%

\end{document}